\documentclass[journal]{IEEEtran}
\ifCLASSINFOpdf
\else
\fi

\usepackage{bbm}
\usepackage{verbatim}
\usepackage{graphicx}
\usepackage{cite}
\usepackage{url}
\usepackage[cmex10]{amsmath}
\usepackage{amssymb}    
\usepackage{amsmath}
\usepackage{amsthm}
\usepackage{algorithm, algorithmicx, algpseudocode}
\usepackage[caption=false,font=footnotesize]{subfig}
\usepackage{color}
\usepackage{cite}
\usepackage{epstopdf}
\usepackage{calc}
\usepackage{array}
\usepackage{stfloats}
\usepackage{siunitx}
\usepackage{mathtools}
\usepackage{optidef}
\usepackage{gensymb}
\usepackage[printonlyused]{acronym}
\usepackage{xcolor}
\usepackage{cleveref}
\usepackage{multirow}
\usepackage{xcolor}
\usepackage{flushend}

\acrodef{mmW}{millimeter-wave}
\acrodef{ULA}{uniform linear array}
\acrodef{sub-THz}{sub-teraherz}
\acrodef{LWA}{Leaky wave antenna}
\acrodef{BS}{base station}
\acrodef{UE}{user equipment}
\acrodef{SOTA}{state of the art}
\acrodef{AoA}{angle of arrival}
\acrodef{AoD}{angle of departure}
\acrodef{AWV}{antenna weight vector}
\acrodef{ADC}{analog-to-digital converter}
\acrodef{AGC}{automatic gain control}
\acrodef{AP}{access point}
\acrodef{BB}{baseband}
\acrodef{RSRP}{reference signal received power}
\acrodef{CSI}{channel state information}
\acrodef{COTS}{commercial-off-the-shelf}
\acrodef{PAA}{phased antenna array}
\acrodef{TTD}{true-time-delay}
\acrodef{LoS}{line-of-sight}
\acrodef{NLoS}{non-line-of-sight}
\acrodef{IA}{initial access}
\acrodef{DFT}{discrete Fourier transform}
\acrodef{UDN}{ultra-dense networks}
\acrodef{RF}{radio frequency}
\acrodef{MPC}{multipath component}
\acrodef{BF}{beamforming}
\acrodef{SNR}{signal-to-noise ratio}
\acrodef{SINR}{signal-to-interference-plus-noise ratio}
\acrodef{OFDM}{orthogonal frequency-division multiplexing}
\acrodef{ML}{maximum likelihood}
\acrodef{DSP}{digital signal processing}
\acrodef{LUT}{lookup table}
\acrodef{MIMO}{multiple-input multiple-output}
\acrodef{IC}{integrated circuits}
\acrodef{PS}{phase shifter}
\acrodef{DAC}{digital-to-analog converter}
\acrodef{EVM}{error vector magnitude}
\acrodef{CP}{cyclic prefix}
\acrodef{FPGA}{field programmable gate arrays}
\acrodef{MSE}{mean squared error}
\acrodef{RMSE}{root mean square error}
\acrodef{MMSE}{minimum mean square error}
\acrodef{VTC}{voltage-to-time converter}
\acrodef{TDC}{time-to-digital converter}
\acrodef{CMOS}{complementary metal–oxide–semiconductor}
\acrodef{MUX}{multiplexer}
\acrodef{CLK}{clock}
\acrodef{PI}{phase interpolator}
\acrodef{FoM}{figure-of-Merit}
\acrodef{HI}{hardware impairment}
\acrodef{CS}{compressive sensing}
\acrodef{RST}{reset}
\acrodef{PM}{phase margin}
\acrodef{SCA}{Switched-Capacitor Arrays}
\acrodef{OTA}{operational transconductance amplifier}

\DeclareMathOperator*{\argmax}{argmax}

\DeclarePairedDelimiter{\ceil}{\lceil}{\rceil}
\DeclarePairedDelimiter{\floor}{\lfloor}{\rfloor}
\DeclarePairedDelimiter\abs{\lvert}{\rvert}

\newcommand\numberthis{\addtocounter{equation}{1}\tag{\theequation}}

\newcommand{\BW}[0]{\mathrm{BW}}

\newcommand{\sigmaN}[0]{\sigma_{\text{N}}}
\newcommand{\SNR}[0]{\mathrm{SNR}}
\newcommand{\fc}[0]{f_{\text{c}}}
\newcommand{\fs}[0]{f_{\text{s}}}
\newcommand{\Ts}[0]{T_{\text{s}}}
\newcommand{\Tcmax}[0]{T_{\text{C-max}}}
\newcommand{\PSCA}[0]{P_{\text{SCA}}}
\newcommand{\PSCAo}[0]{P_{\text{OTAo}}}

\newcommand{\PAGC}[0]{P_{\text{AGC}}}

\newcommand{\PTINW}[0]{P_{\text{TINW}}}
\newcommand{\Csw}[0]{C_{\text{sw}}}
\newcommand{\Cint}[0]{C_{\text{int}}}

\newcommand{\VDD}[0]{\text{VDD}}
\newcommand{\PDESo}[0]{P_{\text{DESo}}}
\newcommand{\PDeSer}[0]{P_{\text{DeSer}}}
\newcommand{\T}[0]{\text{T}}
\newcommand{\R}[0]{\text{R}}
\newcommand{\A}[0]{\text{A}}
\newcommand{\D}[0]{\text{D}}

\newcommand{\I}[0]{\text{I}}
\newcommand{\hermitian}[0]{\text{H}}
\newcommand{\transpose}[0]{\text{T}}
\newcommand{\tot}[0]{\text{tot}}

\begin{document}
%
\title{True-Time-Delay Arrays for Fast Beam Training\\
in Wideband Millimeter-Wave Systems}
%
%
%

\author{Veljko~Boljanovic,~\IEEEmembership{Student~Member,~IEEE,}
        Han~Yan,~\IEEEmembership{Student~Member,~IEEE,}
        Chung-Ching~Lin,~\IEEEmembership{Student~Member,~IEEE,}
        Soumen~Mohapatra,~\IEEEmembership{Student~Member,~IEEE,}
        Deukhyoun~Heo,~\IEEEmembership{Senior~Member,~IEEE,}
        Subhanshu~Gupta,~\IEEEmembership{Senior~Member,~IEEE,}
        and~Danijela~Cabric,~\IEEEmembership{Senior~Member,~IEEE}
\thanks{This work was supported in part by NSF under grants 1718742, 1705026, and 1944688. This work was also supported in part by the ComSenTer and CONIX Research Centers, two of six centers in JUMP, a Semiconductor Research Corporation (SRC) program sponsored by DARPA.}%
\thanks{Veljko Boljanovic, Han Yan, and Danijela Cabric are with the Department of Electrical and Computer Engineering, University of California, Los Angeles, Los Angeles, CA 90095 USA (e-mail:
vboljanovic@ucla.edu, yhaddint@ucla.edu, danijela@ee.ucla.edu).}
\thanks{Chung-Ching Lin, Soumen Mohapatra, Deukhyoun Heo, and Subhanshu Gupta are with the School of Electrical Engineering and Computer Science, Washington State University, Pullman, WA 99164 USA (e-mail: chung-ching.lin@wsu.edu, soumen.mohapatra@wsu.edu, dheo@wsu.edu, sgupta@eecs.wsu.edu).}
}

%
%

\markboth{}%
{}
%



\maketitle

\begin{abstract}
The best beam steering directions are estimated through beam training, which is one of the most important and challenging tasks in millimeter-wave and sub-terahertz communications. Novel array architectures and signal processing techniques are required to avoid prohibitive beam training overhead associated with large antenna arrays and narrow beams. In this work, we leverage recent developments in \ac{TTD} arrays with large delay-bandwidth products to accelerate beam training using frequency-dependent probing beams. We propose and study two \ac{TTD} architecture candidates, including analog and hybrid analog-digital arrays, that can facilitate beam training with only one wideband pilot. We also propose a suitable algorithm that requires a single pilot to achieve high-accuracy estimation of angle of arrival. The proposed array architectures are compared in terms of beam training requirements and performance, robustness to practical hardware impairments, and power consumption. The findings suggest that the analog and hybrid \ac{TTD} arrays achieve a sub-degree beam alignment precision with 66\% and 25\% lower power consumption than a fully digital array, respectively. Our results yield important design trade-offs among the basic system parameters, power consumption, and accuracy of angle of arrival estimation in fast \ac{TTD} beam training.

\end{abstract}


%
\IEEEpeerreviewmaketitle

%
%

\section{Introduction}
\label{sec:introduction}
%
%
%
%
\IEEEPARstart{A}{bundant} spectrum at \ac{mmW} frequencies is seen as the key resource for providing high data rates in the fifth generation of cellular systems \cite{Andrews:5G}. However, the use of \ac{mmW} communication bands comes at the cost of less favorable propagation conditions \cite{Rappaport:propagation}. Both the \ac{BS} and \ac{UE} are required to use large antenna arrays to achieve high \ac{BF} gain and compensate for severe propagation loss. Beam pointing directions are estimated through \textit{beam training}, a procedure that identifies the \ac{AoA} and \ac{AoD} of the dominant propagation path in the wireless channel. Apart from aligning the beams for data communication, knowledge of the \ac{AoA} and \ac{AoD} is of utmost importance for other applications in practical \ac{mmW} systems, including interference nulling and localization \cite{7426565}.


The existing mmW systems utilize analog array architecture with a single transceiver \ac{RF}-chain at both the \ac{BS} and \ac{UE} due to its power efficiency. Such arrays are refereed to as phased arrays since they use adjustable phase shifters to allow coherent signal steering/combining in a desired direction. The existing beam training schemes with phased arrays include various types of extensive beam sweeping, where beams with different pointing directions are synthesised to probe the channel sequentially in order to find the \ac{AoD} and \ac{AoA} \cite{Hosoya:wifi2014, Jeong:sweeping2015, Kim:fast, Zhou:efficient}.
The required number of probing beams linearly scales with the number of antenna elements in the array, which directly translates into beam training overhead and latency. Hence, conventionally used beam sweeping faces scalability challenge in higher \ac{mmW} frequency bands, where more antenna elements will be used to achieve the required \ac{BF} gain.

Previous work that addresses the beam training problem can be divided into two categories. The first category intends to reduce the required number of probing beams. Specifically, the number scales logarithmically with the array size when advanced signal processing techniques that exploit the sparsity of \ac{mmW} channel are used \cite{Zhang:beam_alloc, Yan:compressive, kinget_CS_AOA}. Further, various side-information, e.g., location information and out-of-band measurements \cite{Ali:OOB}, can also be used to reduce the required number of probing beams. The second category aims to enhance the simultaneous channel probing capability by using advanced hardware design \cite{Desai:initial, Barati:directional, Barati:initial, Kalia:bf, Yeh:hybrid, Blandino:hybrid, Karl:lwa, Ghasempour:lwa}. These approaches are more robust when the channel sparsity and side information are not available.
Fully digital array architectures, with a dedicated \ac{RF}-chain per each antenna element, offer the highest flexibility and capability of channel probing. From the signal processing perspective, signals from all antenna branches can be steered/combined to simultaneously probe all angular directions for fast \ac{AoD}/\ac{AoA} estimation \cite{Desai:initial, Barati:directional, Barati:initial}.
%
Fully-connected or sub-band based hybrid arrays are another way to enhance simultaneous probing of the channel \cite{Kalia:bf, Yeh:hybrid, Blandino:hybrid}. They can probe multiple directions simultaneously and the flexibility increases linearly with the number of RF-chains that control phase shifter based analog front-end \cite{Desai:initial}. The probing capability of hybrid arrays can be further enhanced by associating probing beams with different frequencies using spatio-spectral \ac{BF} \cite{Kalia:bf}.
\ac{LWA} can scan all angular directions simultaneously by using different frequency resources since the pointing directions of the beams are frequency-dependent \cite{Karl:lwa, Ghasempour:lwa}.
However, the existing \ac{LWA} technique requires access to THz spectrum for adequate frequency dispersive beam steering.

\ac{TTD} arrays are another appealing, yet insufficiently investigated alternative for fast \ac{mmW} beam training. Due to time delaying of the signal in each antenna branch, \ac{TTD} arrays have frequency-dependent probing beams, which can be exploited to enhance the channel probing capability. Further, the frequency-dependent beams can be fully controlled by adjusting the delay introduced in \ac{TTD} circuits \cite{9048885}. Early implementations relied on delay lines in all antenna branches \cite{Chu:TTD}, but this approach suffered from low scalability in terms of required area and power efficiency when the array size becomes large. Further, limited delay range at \ac{RF} is insufficient to achieve frequency dispersive beam training as proposed in this work. Recent advancement in TTD arrays with baseband delay elements and large delay range-to-resolution ratios \cite{ghaderi2019, ghaderi2020}, improved the scalability and thus enabled the realization of fast beam training schemes with large arrays.



In this paper, we extend our previous work \cite{Boljanovic:TTD} and present the design of baseband \ac{TTD} array architectures for \ac{mmW} beam training. To the best of our knowledge, this is the first work that comprehensively study the system and circuit aspects of TTD based \ac{mmW} beam training with dispersive channel probing. The key contributions are summarized as follows:

\begin{itemize}

\item We propose two \ac{TTD} architecture candidates with baseband delay elements for fast beam training: 1) analog \ac{TTD} architecture, where signal delaying is done in analog baseband domain; 2) hybrid analog-digital \ac{TTD} architecture, where signal delaying is done both in analog and digital domains.

\item We propose a power measurement based beam training scheme for TTD arrays that requires only one training pilot. In particular, we design frequency-dependent probing beams robust to frequency-selective \ac{mmW} channels and a \ac{DSP} algorithm for high-accuracy angle estimation. We numerically evaluate the performance of the proposed algorithm in a practical multipath fading channel.

\item We study the required \ac{TTD} hardware specifications of both array architectures and we explain how hardware constraints affect the beam training performance. We also quantitatively study the beam training performance under practical hardware impairments of TTD array circuits, including phase errors, delay compensation errors, and \ac{ADC} quantization errors. We include the performance of a fully digital array as the benchmark.

\item We model and estimate the power consumption of the proposed \ac{TTD} array architectures in the beam training framework. We investigate how power consumption scales with the key system parameters, including the bandwidth and array size, which provides an insight into the beam training design in future mmW/sub-THz systems. Power consumption of the fully digital array is included as the benchmark.
\end{itemize}

The rest of the paper is organized as follows. In \Cref{sec:ttd_beam_training}, we introduce the two \ac{TTD} architectures and benchmark fully digital array. \Cref{sec:system_and_alg} introduces a wideband system model and it describes the beam training codebook and \ac{DSP} algorithm design. In \Cref{sec:perf}, we explain the baseband implementation of \ac{TTD} elements and compare the considered architectures in terms of the beam training performance under practical hardware impairments. Power consumption of all three considered architectures is modeled and evaluated in \Cref{sec:poweranal}. \Cref{sec:concl} concludes the paper.


%
%
\section{\ac{TTD} Array Architectures for Beam Training}
\label{sec:ttd_beam_training}
The realization and performance of \ac{TTD} beam training schemes heavily depends on the underlying \ac{TTD} hardware. The design of a fast high performance beam training scheme imposes a challenging delay range requirement on \ac{TTD} circuits, which raises the question of a \textit{beam-training-efficient} \ac{TTD} array architecture. In this work, the efficiency depends the number of pilots used in beam training, angle estimation accuracy, and array power consumption. To address this question, we propose and extensively compare two uniform linear array architectures with baseband \ac{TTD} elements, including analog and hybrid analog-digital arrays. We include a fully digital array architecture in the comparison as the benchmark due to its known flexibility and high beam training performance. All three considered array architectures are described in the reminder of this section.


\begin{figure}[t]
    \begin{center}
        \includegraphics[width=0.49\textwidth]{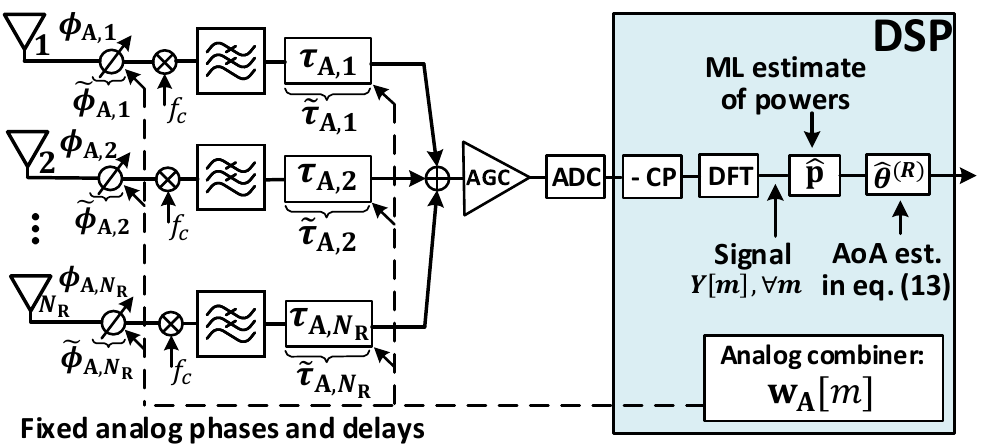}
    \end{center}
    \vspace{-4mm}
    \caption{Architecture of analog TTD array with uniform delay spacing $\Delta\tau$ and phase spacing $\Delta\phi$ between antennas. The design of combiners and \ac{DSP} algorithm is explained in \Cref{sec:algorithm}.}
    \vspace{-4mm}
    \label{fig:analog}
\end{figure}

An analog uniform linear \ac{TTD} array with a single \ac{RF}-chain and $N_{\R}$ antennas is presented in Fig.~\ref{fig:analog}. The $n$-th antenna branch has an analog phase shifter with the phase tap $\phi_{\A,n} = (n-1)\Delta\phi$ and a analog baseband \ac{TTD} element with the delay tap $\tau_{\A,n} = (n-1)\Delta\tau$, where $\Delta\phi$ and $\Delta\tau$ represent the phase and delay spacing between neighboring branches, respectively. Due to the hardware errors in practical phase shifters and \ac{TTD} elements, the phase and delay taps can be distorted. In all antenna branches, we model the time-invariant distorted taps as independent zero-mean Gaussian random variables $\tilde{\phi}_{\A,n}\sim\mathcal{N}\left(\phi_{\A,n},\sigma_{\text{P}}^2\right)$ and $\tilde{\tau}_{\A,n}\sim\mathcal{N}\left(\tau_{\A,n},\sigma_{\text{T}}^2\right)$, respectively. For a specific delay spacing $\Delta\tau$, \ac{TTD} frequency-dependent \ac{AWV} results in a fixed beam training codebook, where different frequency components of the signal are hard-coded in different angular directions. The frequency-flat phase shifters increase the flexibility by enabling codebook rotations and different frequency-to-angle mapping. The maximum delay in the $N_{\R}$-th antenna branch is $\tau_{\A,N_{\R}} = (N_{\R}-1)\Delta\tau$, which becomes an implementation bottleneck for large antenna arrays. The state-of-the-art \ac{TTD} delay range is in the order of \SI{15}{\nano\second} \cite{ghaderi2019}, which can be insufficient for wideband beam training with a moderate number of antenna elements $N_{\R}$, e.g., $N_{\R} = 32$, as we previously discussed in \cite{Boljanovic:TTD}.

\begin{figure*}[t]
    \begin{center}
        \includegraphics[width=1\textwidth]{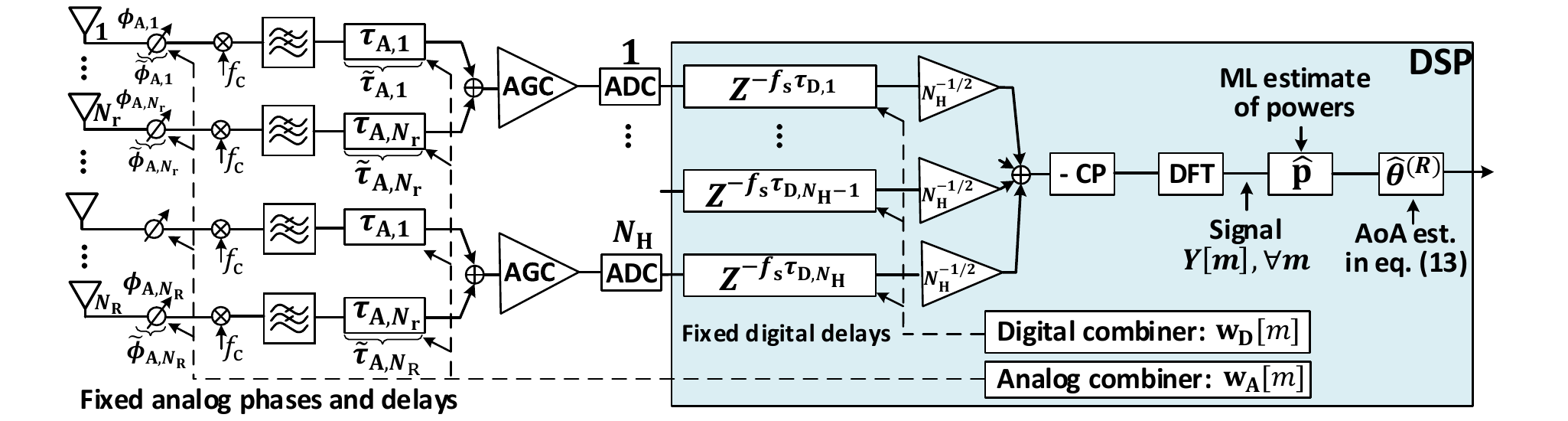}
    \end{center}
    \vspace{-4mm}
    \caption{Architecture of hybrid analog-digital TTD array with uniform delay spacing $\Delta\tau$ and phase spacing $\Delta\phi$ between antennas. The design of combiners and \ac{DSP} algorithm is explained in \Cref{sec:algorithm}.}
    \vspace{-4mm}
    \label{fig:hybrid}
\end{figure*}
\begin{figure*}[t]
    \begin{center}
        \includegraphics[width=1\textwidth]{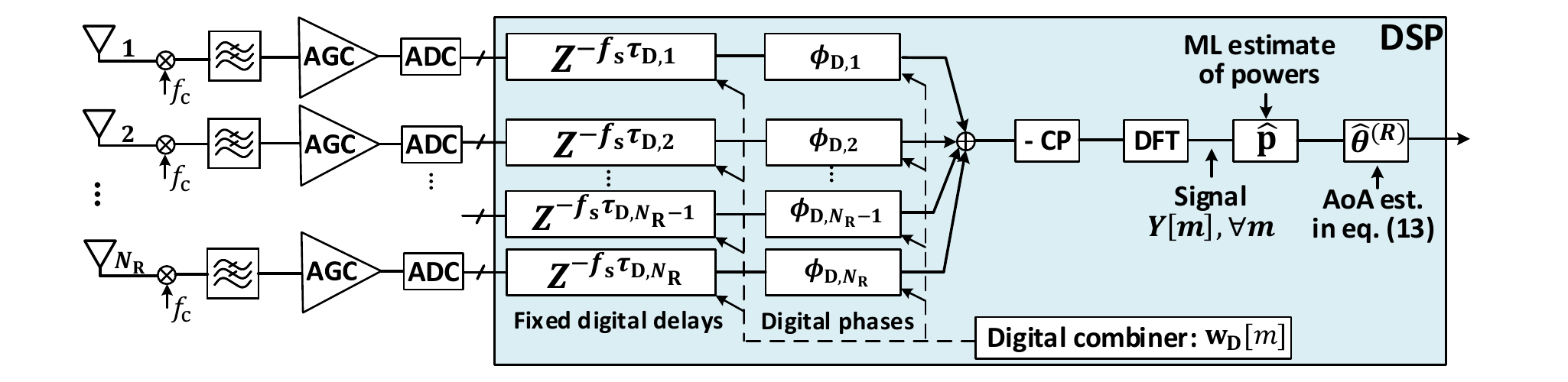}
    \end{center}
    \vspace{-4mm}
    \caption{Architecture of the benchmark fully digital array that emulates a \ac{TTD} architecture by introducing digital delays. The design of combiners and \ac{DSP} algorithm is explained in \Cref{sec:algorithm}.}
    \vspace{-4mm}
    \label{fig:digital}
\end{figure*}
To alleviate the delay range requirement and improve the scalability of analog \ac{TTD} arrays, we introduce a hybrid analog-digital architecture with $N_{\text{H}}$ sub-arrays, each controlled by one distinct \ac{RF}-chain, as illustrated in Fig.~\ref{fig:hybrid}. The hybrid array uses a combination of analog and digital signal delaying, where first all the sub-arrays of $N_{\text{r}}$ antennas introduce the same delays $\tau_{\A,n^{\prime}}=(n^{\prime}-1)\Delta\tau,~n^{\prime}=1,...,N_{\text{r}}$, in the analog domain. The relative delay difference among antennas is compensated in the digital domain by introducing the fixed digital taps $\tau_{\D,h}= (h-1)N_{\text{r}}\Delta\tau,~h=1,...,N_{\text{H}}$, i.e., digital delays $\fs \tau_{\D,h}$, where $\fs$ is the sampling frequency. As in the analog \ac{TTD} array, the distorted phase taps $\tilde{\phi}_{\A,n},~n=1,...,N_{\R}$, and delay taps $\tilde{\tau}_{\A,n},~n=1,...,N_{\R}$, are modeled as independent Gaussian random variables.

A fully digital array, used as the benchmark, is illustrated in Fig.~\ref{fig:digital}. The digital array can emulate a \ac{TTD} array through \ac{DSP} by using the fixed digital taps $\tau_{\D,n}=(n-1)\Delta\tau,~n=1,...,N_{\R}$, i.e., digital delays $\fs \tau_{\D,n}$ in the corresponding antenna branches. The ability to control the digital phases $\phi_{\D,n},~n=1,...,N_{\R}$, in \ac{DSP}, allows the signal frequency components to be independently steered/combined in any angular direction, which provides high flexibility in the beam training design. The digital array does not have the phase shifters and \ac{TTD} elements before the \ac{ADC}s, thus we assume it is insensitive to hardware errors. However, each antenna element has a dedicated \ac{RF}-chain, which significantly affects the array power efficiency, as discussed later in Section~\ref{sec:poweranal}.


In the next section, we explain how $\Delta\tau$ and $\Delta\phi$ are set up in all three architecture to obtain a beam training codebook robust to frequency-selective channels. We also introduce a \ac{DSP} algorithm that exploits this codebook. Based on the designed $\Delta\tau$, \Cref{sec:perf} discusses the requirements in \ac{TTD} hardware implementation and impact of hardware impairments on the beam training performance. Accounting for the designed $\Delta\tau$ and proposed baseband \ac{TTD} implementation, we compare the three architectures in terms of power consumption in \Cref{sec:poweranal}.

%
%

\section{TTD Beam Training Algorithm Design}
\label{sec:system_and_alg}

\begin{figure*}[t]
    \begin{center}
        \includegraphics[width=1\textwidth]{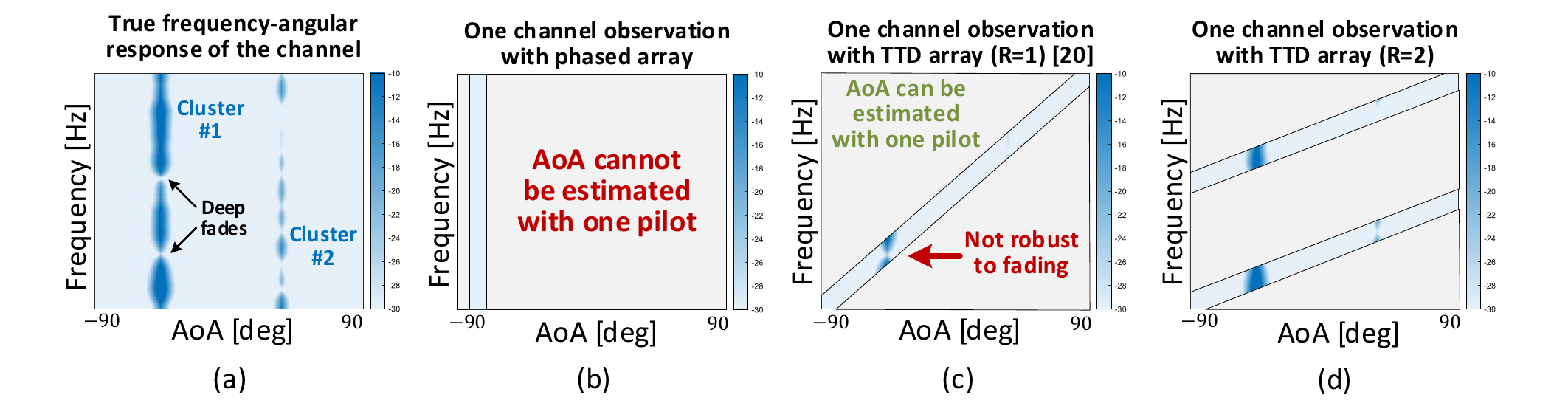}
    \end{center}
    \vspace{-6mm}
    \caption{Beam training in clustered frequency-selective multipath channel: (a) An example of frequency-selective channel with two multipath clusters. Frequency-selectivity comes from intra- and inter-cluster delay spreads. (b) Channel observation of a phased array when only one pilot is used. Beam sweeping is necessary to cover all angles in the range $(-\pi/2, \pi/2)$. (c) Channel observation of a \ac{TTD} array when only one pilot is used. Frequency components (subcarriers) are mapped into different angles to simultaneously probe the range $(-\pi/2, \pi/2)$. The angle estimation may fail in frequency-selective channels. (d) Enhanced \ac{TTD} codebook with frequency diversity order $R=2$.} 
    \vspace{-4mm}
    \label{fig:frequency_robustness}
\end{figure*}

In this section, we describe a \ac{DSP} algorithm which achieves a high angle estimation accuracy using only one pilot symbol in a clustered frequency-selective multipath channel. An example of such channel with one strong and one weak cluster, as seen by the receiver, is provided in Fig.~\ref{fig:frequency_robustness}(a). Conventional phased arrays cannot estimate the \ac{AoA} of the dominant cluster with one training pilot, and thus they require beam sweeping, as illustrated in Fig.~\ref{fig:frequency_robustness}(b). On the other hand, \ac{TTD} arrays are capable of estimating the \ac{AoA} fast, but the corresponding \ac{DSP} algorithm must include the design of a suitable \ac{TTD} beam training codebook.

\subsection{System Model}
\label{sec:system}

We consider downlink beam training between the \ac{BS} and \ac{UE}, where the \ac{CP} based \ac{OFDM} waveform is used as a training pilot. The carrier frequency, bandwidth, and number of subcarriers are denoted as $\fc$, $\BW$, and $M_{\tot}$, respectively. The power-normalized training pilot uses $M$ subcarriers from the predefined set $\mathcal{M}$, all loaded with the same binary phase shift keying modulated symbol. Both the \ac{BS} and \ac{UE} have half-wavelength spaced uniform linear arrays with $N_{\T}$ and $N_{\R}$ antennas, respectively. We assume that AoD at the \ac{BS} has already been estimated so that BS uses a fixed frequency-flat beam defined by a precoder vector $\mathbf{v}\in\mathbb{C}^{N_{\T}}$. The \ac{UE} is equipped with a \ac{TTD} array and it performs beam training to estimate \ac{AoA} $\hat{\theta}^{(\R)}$. Thus, the received signal $Y[m]$ at the $m$-th subcarrier is
\begin{equation}
    Y[m] = \mathbf{w}^{\hermitian}[m] \mathbf{H}[k] \mathbf{v} + \mathbf{w}^{\hermitian}[m]\mathbf{n}[m],~ m\in \mathcal{M},
\label{eq:received_signal_sc}
\end{equation}
where $\mathbf{H}[k]\in \mathbb{C}^{N_{\R} \times N_{\T}}$ is the channel matrix of the $k$-th out of $K_c$ sub-bands in a frequency-selective channel and $\mathbf{n}\sim \mathcal{CN}\left(0, \sigma_{\text{N}}^2 \mathbf{I}_{N_{\R}} \right)$ is white Gaussian noise. Each sub-band contains multiple adjacent sub-carriers, and the relationship between the sub-band index $k$ and subcarrier index $m$ is given as $k=\ceil*{(mK_c)/M_{\tot}}$, where $\ceil{x}$ rounds $x$ to the nearest greater integer.  We assume that all subcarriers within the same frequency sub-band $k$ experience the same channel $\mathbf{H}[k]$. The \ac{TTD} combiner $\mathbf{w}[m]\in \mathbb{C}^{N_{\R}}$ of the $m$-th subcarrier can be decomposed as a Hadamard product $\mathbf{w}[m] = \mathbf{w}_{\A}[m] \odot \mathbf{w}_{\D}[m]$, where the analog combiner $\mathbf{w}_{\A}[m]\in\mathbb{C}^{N_{\R}}$ and digital combiner $\mathbf{w}_{\D}[m]\in\mathbb{C}^{N_{\R}}$ depend on the underlying \ac{TTD} array architecture. In an analog \ac{TTD} array, $\mathbf{w}_{\D}[m]=\mathbf{1}_{N_{\R}}$ since both the phases $\phi_{\A,n},~\forall n$, and delays $\tau_{\A,n},~\forall n$, are introduced in the analog domain. Similarly, $\mathbf{w}_{\A}[m]=\mathbf{1}_{N_{\R}}$ with a fully digital array, as it is insensitive to hardware impairments and the phases $\phi_{\D,n},~\forall n$, and delays $\tau_{\D,n},~\forall n$, are applied in the digital domain.
In general, the $n$-th elements of $\mathbf{w}_{\A}[m]$ and $\mathbf{w}_{\D}[m]$ are
\begin{align}
    \label{eq:TTD_AWV_A}
    \left[ \mathbf{w}_{\A}[m] \right]_n &= \mathrm{exp}\left[{-j\left(2\pi (f_m-\fc)\tilde{\tau}_{\A, n} + \tilde{\phi}_{\A,n} \right)}\right]\\
    \label{eq:TTD_AWV_D}
    \left[ \mathbf{w}_{\D}[m] \right]_n &= \mathrm{exp}\left[{-j\left(2\pi (f_m-\fc)\tau_{\D,n} + \phi_{\D,n} \right)}\right]
\end{align}
where $f_m =\fc - \BW/2 + (m-1)\BW/(M_{\tot}-1)$.

The expressions (\ref{eq:TTD_AWV_A}) and (\ref{eq:TTD_AWV_D}) indicate that the beam pointing direction depends on the subcarrier frequency, phases, and delays. With a proper configuration of the phase and delay taps in the analog and/or digital domain, it is possible to set up a codebook of combiners that covers all angular directions, as we discuss in the next subsection.

\subsection{DSP Algorithm for Beam Training}
\label{sec:algorithm}

In this subsection, we first present the design of a robust codebook and then describe a \ac{DSP} algorithm for \ac{TTD} arrays \cite{Boljanovic:TTD} that achieves a high resolution in \ac{AoA} estimation.

In \cite{9048885}, we have demonstrated that $D$ spatial directions in the angular range $(-\pi/2, \pi/2)$ can be sounded with a single \ac{OFDM} symbol by mapping one subcarrier per direction, as illustrated in Fig.~\ref{fig:frequency_robustness}(c). We have shown that this can be achieved with an analog \ac{TTD} array by setting the delay spacing to be $\Delta\tau=1/\BW$. The resulting codebook is, however, sensitive to frequency-selective channels since certain subcarriers can experience deep fades and thus miss to detect the incoming signal. The codebook can be enhanced by increasing its frequency diversity order $R$, i.e., by mapping $R$ distinct subcarriers in each probed direction \cite{Boljanovic:TTD}. The benefit of the enhanced codebook is illustrated in Fig.~\ref{fig:frequency_robustness}(d) for $R=2$, where two subcarriers detect the dominant cluster. To increase the diversity, we define $D$ distinct sets $\mathcal{M}_d,~1\leq d \leq D$, of $R$ subcarriers, where each set is associated with a different direction $d,~1\leq d \leq D$. Mathematically, the $R$ subcarriers from the set $\mathcal{M}_d$ have the same combiner $\mathbf{f}_d$, i.e., $\mathbf{w}[m] = \mathbf{f}_d,~\forall m \in \mathcal{M}_d$, where the $n$-th element of $\mathbf{f}_d$ is defined as
\begin{equation}
    [\mathbf{f}_{d}]_n = \mathrm{exp}[-j2\pi (n-1)(d-1-D/2)/D],~~ d\leq D.
    \label{eq:DFT_beams}
\end{equation}

\begin{figure}[t]
    \begin{center}
        \includegraphics[width=0.48\textwidth]{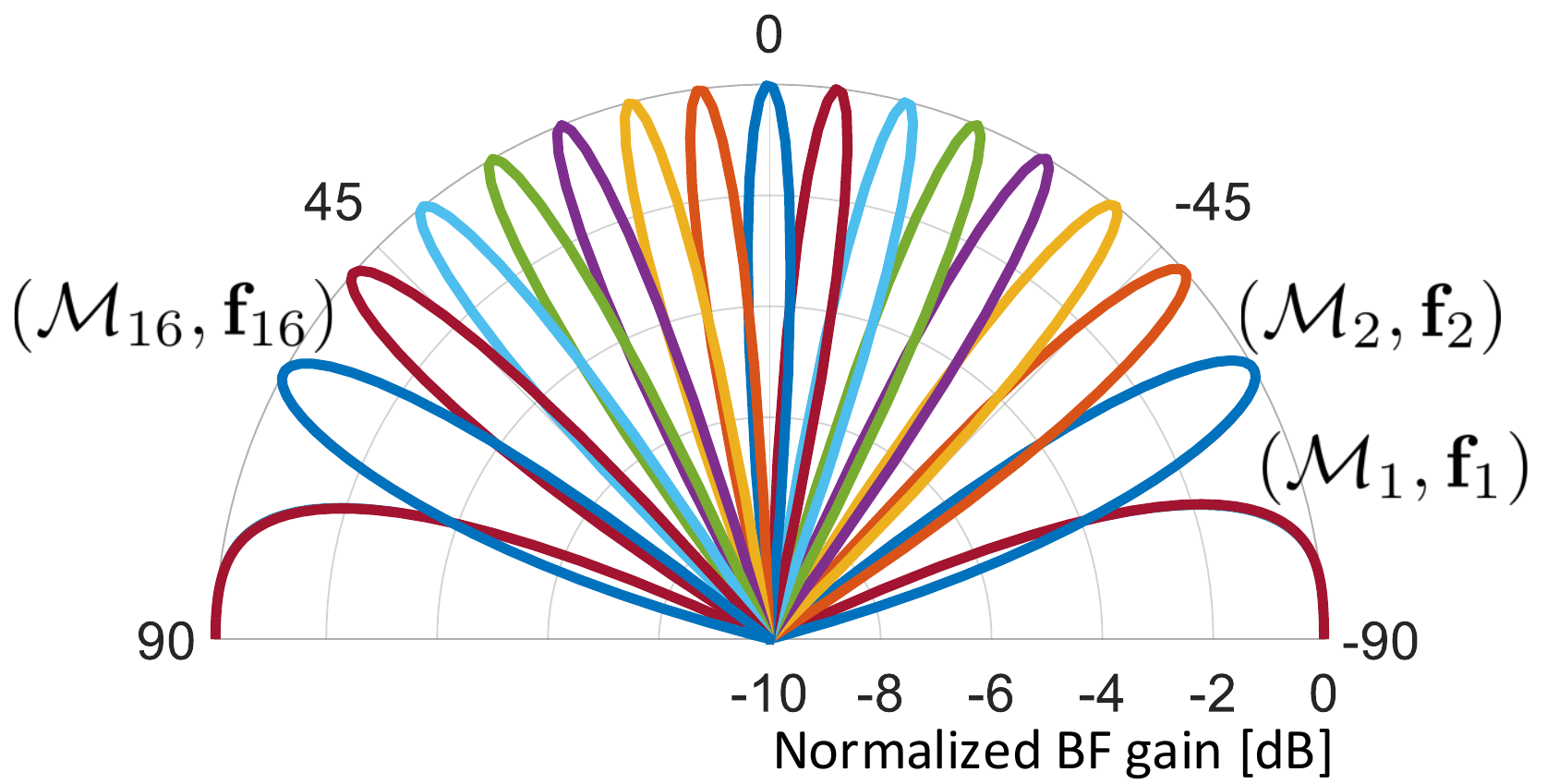}
    \end{center}
    \vspace{-4mm}
    \caption{An example of robust \ac{TTD} codebook for $N_{\R}=16$, $D=16$, and $R=4$. All $D=16$ directions are probed simultaneously. Direction $d,~1\leq d \leq D$, is associated with set of subcarriers $\mathcal{M}_d$ and combiner $\mathbf{f}_d$.}
    \vspace{-4mm}
    \label{fig:beams}
\end{figure}
\noindent The subcarriers in $\mathcal{M}_d,~\forall d$, however, should experience different channels, and thus we choose them uniformly across the bandwidth with the step size larger or equal than the coherence bandwidth (channel sub-band size). This codebook can be created for an analog \ac{TTD} array by setting the $n$-th phase and delay taps as follows
\begin{align}
    \label{eq:analog_phi}
    \phi_{\A,n} &= (n-1)[\mathrm{sgn}(\psi)\pi - \psi]\\
    \label{eq:analog_tau}
    \tau_{\A,n} &= (n-1)R/\BW,
\end{align}
where $\psi = \mathrm{mod}(2\pi R (\fc-\BW/2)/\BW + \pi, 2\pi) - \pi$. $\mathrm{sgn}()$ and $\mathrm{mod}()$ are the sign and modulo operators, respectively. The phase taps in (\ref{eq:analog_phi}) ensure that the first set of subcarriers $\mathcal{M}_1$ is mapped into the first probed angle ($-\pi/2$). An example of the resulting codebook with $N_{\R}=16$, $D=16$, and $R=4$ is provided in Fig.~\ref{fig:beams}. Note that the same enhanced codebook can be created for the hybrid \ac{TTD} or fully digital array without the need to implement a fractional \ac{ADC} sampling since $\Delta\tau$ is proportional to the Nyquist sampling period, i.e., $\Delta\tau = R/\BW$. Analog and digital delay taps of the hybrid array introduced in \Cref{sec:ttd_beam_training}, can be expressed with respect to the indices of all antenna elements in the array $n=1,..,N_{\R}$, as $\tau_{\A,n} = \left(n-1-\floor{(n-1)/N_{\text{r}}}N_{\text{r}}\right)\Delta\tau$, and $\tau_{\D,n} = \floor{(n-1)/N_{\text{r}}}N_{\text{r}}\Delta\tau$, respectively. The operator $\floor{x}$ rounds $x$ to the nearest lower integer. Thus, the hybrid \ac{TTD} array can create the enhanced codebook by setting the $n$-th taps of its analog and digital combiners in the following way
\begin{align}
    \label{eq:hybrid_analog_phi}
    \phi_{\A,n} &= (n-1)[\mathrm{sgn}(\psi)\pi - \psi],\\
    \label{eq:hybrid_analog_tau}
    \tau_{\A,n} &= \left(n-1-\floor{(n-1)/N_{\text{r}}}N_{\text{r}}\right)R/\BW,\\
    \label{eq:hybrid_digital_tau}
    \tau_{\D,n} &= \floor{(n-1)/N_{\text{r}}}N_{\text{r}}R/\BW,
\end{align}
where $\psi$ is defined as earlier. The result in (\ref{eq:hybrid_digital_tau}) suggests that the $h$-th sub-array needs to introduce a digital delay of $2(h-1)N_{\text{r}}R$ time samples, assuming the Nyquist sampling frequency $\fs=2\BW$. The considered hybrid array in Fig.~\ref{fig:hybrid} does not apply the phase changes in the digital domain. The digital array can create the enhanced codebook by using the following digital taps
\begin{align}
    \label{eq:digital_phi}
    \phi_{\D,n} &= (n-1)\Delta\phi,~~\Delta\phi \in \mathbb{R}\\
    \label{eq:digital_tau}
    \tau_{\D,n} &= (n-1)R/\BW.
\end{align}
The phase tap in (\ref{eq:digital_phi}) implies that the digital array can leverage the \ac{DSP} to introduce any number of phase spacings $\Delta\phi$. With $\fs=2\BW$, the $n$-th antenna branch will introduce the digital delay of $2(n-1)R$ time samples according to (\ref{eq:digital_tau}).

The phase and delay taps required for the design of a robust codebook are summarized in \Cref{tab:awv_settings} for all three arrays.

\begin{table}[h]
\vspace{-3mm}
\centering
\caption{Phase and delay tap settings for robust codebook design} 
\label{tab:awv_settings}
\vspace{-2mm}
{   
    \footnotesize { 
    \begin{tabular}{|c|c||c|c|c|c|} 
    \hline
        Array arch. & $\mathbf{w}[m]$ & $\phi_{\A,n}$ & $\tau_{\A,n}$ & $\phi_{\D,n}$ & $\tau_{\D,n}$\\
        \hline
        \hline
        Analog TTD  & $\mathbf{w}_{\A}[m]$ & (\ref{eq:analog_phi}) & (\ref{eq:analog_tau}) & N/A & N/A \\
        \hline
        Hybrid TTD & $\mathbf{w}_{\A}[m]\odot\mathbf{w}_{\D}[m]$  & (\ref{eq:hybrid_analog_phi}) & (\ref{eq:hybrid_analog_tau}) & N/A & (\ref{eq:hybrid_digital_tau}) \\
        \hline
        Digital & $\mathbf{w}_{\D}[m]$ & N/A & N/A & (\ref{eq:digital_phi}) & (\ref{eq:digital_tau}) \\
        \hline
    \end{tabular} \\
}
}
\vspace{1mm}
\end{table}
We note that the analog and hybrid \ac{TTD} architectures have the same limited flexibility of receive combining in beam training. Namely, once their corresponding analog combiners $\mathbf{w}_{\A}[m],~m\in\mathcal{M}$, and digital combiners $\mathbf{w}_{\D}[m],~m\in\mathcal{M}$ are set up, they cannot be further changed or manipulated in \ac{DSP}. In both architectures, this happens because the signals from different antenna branches are completely or partially combined before passing through \ac{ADC}s. Thus, the inability to rotate the combiners limits the number of sounded directions to $D$ in both arrays. The diversity order $R$ is also limited, but not necessarily the same in both arrays, as discussed later in the paper. On the other hand, the digital array can exploit digitized signals in all antenna branches and combine them from many different directions in \ac{DSP} by changing the phases $\phi_{D,n},~\forall n$. Different phases $\phi_{D,n}$ introduces angular shifts of the entire codebook, and enable scanning more angles and/or higher diversity.

We use the designed beam training codebook to develop a non-coherent power-based \ac{DSP} angle estimation algorithm. Non-coherent algorithms are preferred in \ac{mmW} beam training as they avoid complex joint synchronization and beam training receiver processing.

Since the subcarriers from $\mathcal{M}_d,~\forall d$, experience different channels, we can consider the received signal in all $D$ probed directions as random. In a clustered multipath channel, the vector of expected powers in $D$ directions $\mathbf{p} = \left[p_1, p_2, ..., p_D \right]^{\transpose}$ can be expressed as
\begin{align}
    \label{eq:expected_power_matrix}
    \mathbf{p} = \mathbf{B}\mathbf{g} + N_{\R}\sigmaN^2\mathbf{1},
\end{align}
where $\mathbf{B}\in \mathbb{R}^{D \times Q}$ is a known dictionary obtained by generalizing the \ac{UE} \ac{BF} gains in $Q$ angles $\xi_q,~q=1,...,Q$, for all $D$ combiners.
The $(d,q)$-th element of $\mathbf{B}$ is defined as $[\mathbf{B}]_{d,q} =  \abs{\mathbf{f}_d^{\hermitian}\mathbf{a}_{\R}(\xi_q)}^2$, where $\mathbf{a}_{\T}(\xi_q)$ is the receive spatial response with elements $[\mathbf{a}_{\R}(\xi_q)]_n = N^{-1/2}_{\R} \mathrm{exp}({-j(n-1)\pi\sin({\xi_q)}}),~n=1,...,N_{\R}$.
The vector $\mathbf{g}\in \mathbb{R}^{Q}$ has only one non-zero element. For a detailed derivation of (\ref{eq:expected_power_matrix}), please refer to Appendix~\ref{sec:alg_appendix}.

During beam training, the estimates of $p_d,~\forall d$, are obtained by averaging out the powers of all subcarriers from the corresponding set $\mathcal{M}_d,~\forall d$ \cite{Boljanovic:TTD}. In fact, it can be shown that the sample mean is the \ac{ML} estimator of $p_d,~\forall d$. The vector of power estimates is denoted as $\hat{\mathbf{p}}$, which approximate $\mathbf{p}$ in (\ref{eq:expected_power_matrix}). Based on the power measurement model in (\ref{eq:expected_power_matrix}), \ac{AoA} estimation can be solved based on the \ac{ML} criterion using simple linear algebra operations. The \ac{AoA}  $\theta^{(\R)}$ estimate is obtained by finding the index of the column in $\mathbf{B}$ which has the highest correlation with $\hat{\mathbf{p}}$, which is mathematically expressed as
\begin{equation}
    \hat{\theta}^{(\R)} = \xi_{q^{\star}}, \text{ where } q^{\star} = \argmax_q \frac{\hat{\mathbf{p}}^{\transpose}[\mathbf{B}]_{:,q}}{||[\mathbf{B}]_{:,q}||}.
    \label{eq:angle_estimate1}
\end{equation}
The proposed algorithm can achieve high \ac{AoA} estimation accuracy by increasing $Q$, i.e., the number of the columns in the dictionary matrix $\mathbf{B}$. Although this increases the \ac{DSP} complexity, the proposed beam training scheme can still be performed with a single \ac{OFDM} symbol. For the rest of this paper, we use \ac{RMSE} of \ac{AoA} estimation and power consumption as main metrics for the comparison of the proposed \ac{TTD} architectures. The \ac{AoA} \ac{RMSE} closely describes the beam training performance and it can be directly converted to an alternative metric in other applications, including the spectral efficiency in \ac{mmW} data communication and position error in localization.

%
%

\begin{figure}
  \centering
  \begin{tabular}{@{}c@{}}
    \includegraphics[width=1\columnwidth]{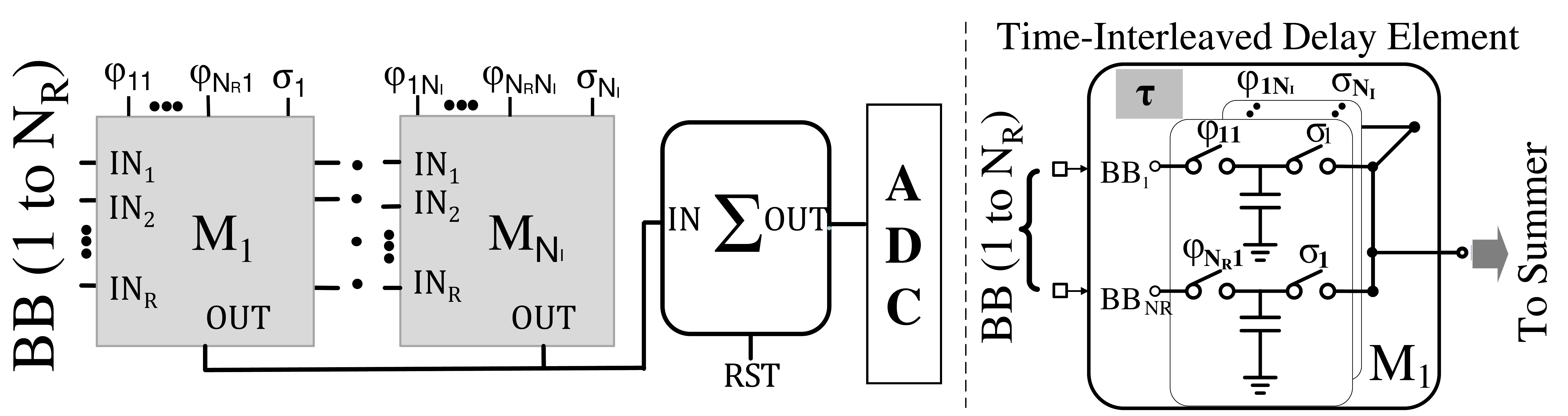} \\[\abovecaptionskip]
    \small (a)
    \label{fig:fig5(a)}
  \end{tabular}
  \begin{tabular}{@{}c@{}}
    \includegraphics[width=1\columnwidth]{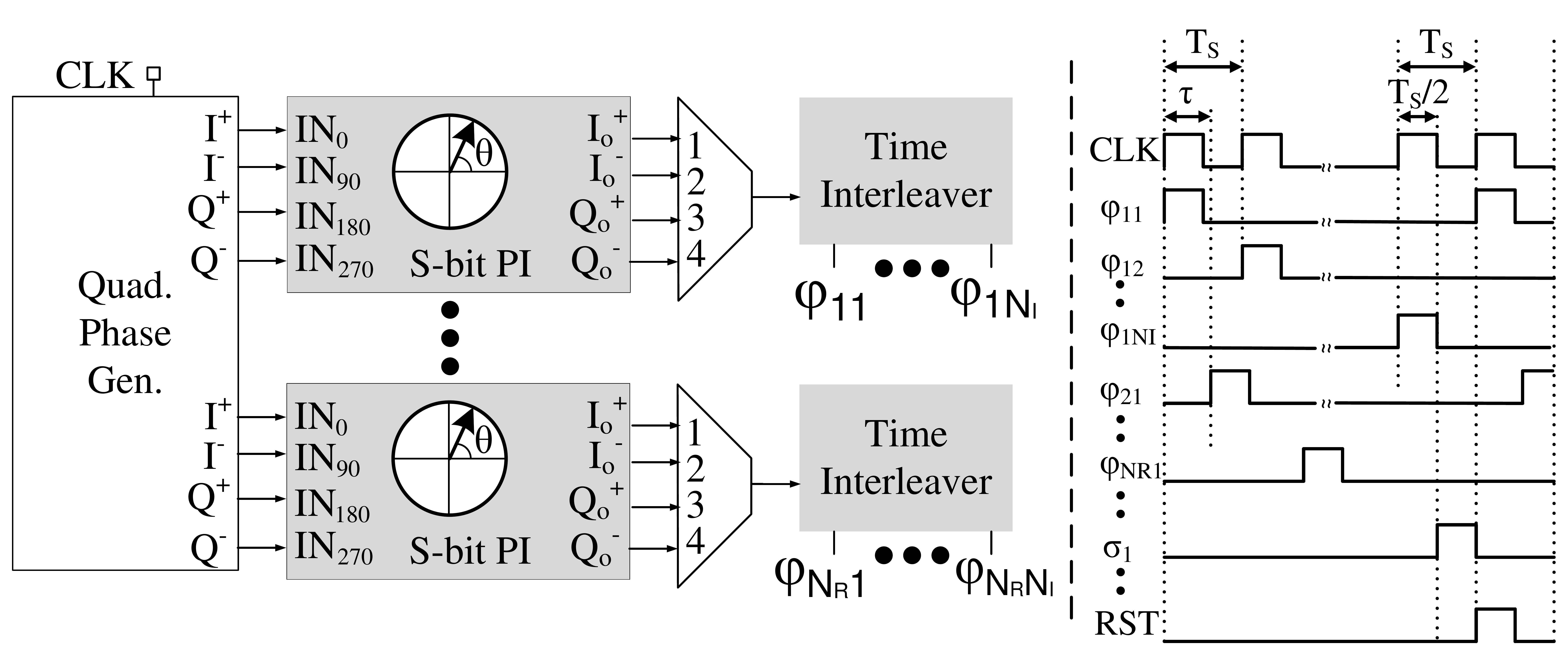} \\[\abovecaptionskip]
    \small (b)
    \label{fig:fig5(b)}
  \end{tabular}
  \caption{(a) Multiply-and-accumulate in discrete-time for TTD \ac{BF} \cite{ghaderi2019,ghaderi2019b} (inset: switched-capacitor adder without the opamp) and (b) time-interleaved clock generation unit (inset: example timing diagram).}
  \label{fig:sca}
\end{figure}

\section{Architecture Performance Analysis}
\label{sec:perf}
In this section, we introduce and compare the baseband implementation of analog \ac{TTD} elements in analog and hybrid \ac{TTD} architectures. Then we study the impact of limited \ac{TTD} delay range in both architectures on beam training performance and we explain the interplay between the number of antenna elements $N_{\R}$, bandwidth $\BW$, and diversity order $R$. We also numerically evaluate the impact of hardware impairments and \ac{ADC} quantization error on the \ac{AoA} estimation accuracy. 


\subsection{Baseband Implementation of Analog \ac{TTD} Front-End}

\noindent While \ac{TTD} array operation is conceptually simple, its physical implementation is non-trivial when targeting large delay range. In general, implementing delays with large range-to-resolution ratios is difficult without severe penalties in linearity, noise, power and area besides increased design complexity. In an array with baseband \ac{TTD} elements, instead of delaying the down-converted and phase shifted signals from the antennas, sampling and digitization, the signals are sampled at different time instants through the \ac{SCA} circuit, resulting in the same digitized value. Thus, the complexity of delaying signals is shifted to the clock path where precise and calibrated delays can be applied in the advanced semiconductor technology nodes. More importantly, a large delay range-to-resolution ratio can be realized easily. The SCA based implementation requires multiple time-interleaved and delay-compensated phases for formation of the beam as shown in Fig.~\ref{fig:sca}(a) and discussed in detail in \cite{ghaderi2019}. In the sampling phase, the input signal from each channel is first sampled (with delayed clocks) on a sampling capacitor (C\textsubscript{S}). After the last sampling phase, the stored charges on each capacitor corresponding to each channel (and each time-interleaved phase) are summed to form the beam. 

The proposed beam-training algorithm requires wider delay ranges with delay offsets that are integer multiples of $\Delta\tau$. This significantly relaxes the design requirement of the SCA and the clock path for TTD-based beam-training. Larger delay-bandwidth products can thus be realized using passive SCA whose performance will not be limited by the opamp feedback factor or time-based circuits as demonstrated in our recent work in \cite{ghaderi2020}. Ongoing research is also investigating use of high-linearity and high-speed ring amplifiers \cite{Hershberg2012} in the \ac{SCA}. 

Fig.~\ref{fig:sca}(b) shows the clock generation circuit. The proposed beam-training just requires a time-interleaver applied to the input clock. The output of the time-interleaver is applied to interleaved multiply-and-accumulate units (MAC) in the SCA (=\textit{N}\textsubscript{I}) and enables the SCA to span the required delay range while meeting the Nyquist BW. The same circuit can be extended for data communication with the only addition being a multi-bit phase interpolator (PI) as described in \cite{ghaderi2019b}. In Fig.~\ref{fig:sca}(b), the external single-phase \ac{CLK} is first fed to a quadrature phase generator circuit. The quadrature outputs (I-, I+, Q-, Q+) of each phase generator are further fed to the \textit{S}-bit PI.  The quadrature output is then applied to a \ac{MUX} which helps in spanning the angular range $\left(-\pi/2, \pi/2 \right)$. An example timing diagram is also shown in Fig.~\ref{fig:sca}(b) with $N_{\R}=4$ and $N_{\I}=7$ for $R=1$ in a hybrid array. A total of 36 phases are shown at the time-interleaver output with a 12.5$\%$ pulse width. 

We further analyze the number of interleaving levels that are required in analog and hybrid TTD arrays. Considering $N_{\I}$ as the interleaving factor in the analog \ac{TTD} array (Fig.~\ref{fig:analog}), the maximum achievable delay compensation $\Tcmax$ is
\begin{align}
    \Tcmax = (N_{\I}-1)\Ts = (N_{\I}-1)/\fs
\end{align}
where $\Ts$ and $\fs$ are the reference clock period and sampling frequency respectively. To cover the entire angular range in beam training, $\Tcmax$ should be equal to $\tau_{\A,N_{\R}}$. Substituting (\ref{eq:analog_tau}) in this equality and solving for $\Ts$ yields  
\begin{equation}
    \Ts = (N_{\R}-1)R/\left((N_{\I}-1)\BW \right)
\end{equation}
Considering a heterodyne receiver architecture and perfect sampled signal reconstruction satisfying the Nyquist condition (i.e., $\Ts\leq 1/(2\BW)$), $N_{\I}$ can be derived to be
\begin{align}
    \label{eq:max_dly_comp}
    N_{\I} \geq 1+2R(N_{\R}-1).
\end{align}
Equation (\ref{eq:max_dly_comp}) can be further applied for hybrid arrays substituting $N_{\R}$ with $N_{\R}$/$N_{\text{H}}$. 

\begin{table}[t]
\vspace{0mm}
\centering
\caption{Analog TTD array complexity with increased diversity $R$.}
\label{tab:ttdspec}
\vspace{-2mm}
{   
    \small
    \begin{tabular}{|c||c||c||c||c||c|} 
    \hline
        \multirow{2}{*}{$R$}  & \multirow{2}{*}{$\Delta\tau$} & $\tau_{\A,N_{\R}}$ & $N_{\I}$ &  $\tau_{\A,N\textsubscript{r}}$ & $N_{\I}$\\ & & Analog & Analog  & Hybrid & Hybrid\\
        \hline
        \hline
        1  &  \SI{0.5}{\nano\second} & \SI{7.5}{\nano\second} & 31 & 1.5~ns & 7  \\
        \hline
        2  &  \SI{1}{\nano\second} & \SI{15}{\nano\second} & 61 & \SI{3}{\nano\second} & 13        \\
        \hline 
        4  &  \SI{2}{\nano\second} & \SI{30}{\nano\second} & 121 & \SI{6}{\nano\second} & 25 \\
        \hline 
    \end{tabular} \\
}
\vspace{1mm}
Assumed parameters are $N_{\R} = 16, f_{\text{CLK}} =$ \SI{4}{\giga\hertz}, $\BW = $ \SI{2}{\giga\hertz}. Hybrid TTD array has four 4-element sub-arrays (\textit{N}\textsubscript{r}=4). 
\end{table}
\vspace{0mm}

Table~\ref{tab:ttdspec} shows an example case study of the required number of interleaving stages in the analog/hybrid TTD array as a function of diversity order and the delay range. This table uses (\ref{eq:max_dly_comp}) with a specific case of 2~GHz bandwidth, 4~GHz sampling frequency, and 16 antenna elements for both the analog and hybrid array presented in Fig.~\ref{fig:analog} and Fig.~\ref{fig:hybrid} respectively.

\subsection{Impact of Limited \ac{TTD} Delay Range on Beam Training}

In this subsection, we assume that the analog and hybrid architectures have \ac{TTD} elements with the same state-of-the-art maximum delay compensation of $\Tcmax = 15$ ns, or equivalently the same interleaving factor $N_{\I}$.

To realize the proposed beam training algorithm, $\tau_{\A,N_{\R}} \leq \Tcmax$ needs to be satisfied for the analog, and $\tau_{\A,N_{\text{r}}} \leq \Tcmax$ for the hybrid \ac{TTD} array. Based on these conditions, it is straightforward to show that the achievable diversity order $R$ is limited as
\begin{align}
    1 \leq R \leq \frac{\Tcmax}{N_{\R}-1} \BW ~~~ \text{and} ~~~  1 \leq R \leq \frac{\Tcmax}{N_{\text{r}}-1} \BW,
    \label{eq:achievable_R}
\end{align}
for the analog and hybrid array, respectively. Note that with $R<1$, the beam training algorithm cannot be realized with a single \ac{OFDM} symbol. The expressions in (\ref{eq:achievable_R}) describe the dependency of $R$ on the basic system parameters $N_{\R}$, $N_{\text{r}}$, and $\BW$. In the remainder of this subsection, we numerically evaluate the interplay among them.

\begin{figure}
    \begin{center}
        \includegraphics[width=0.48\textwidth]{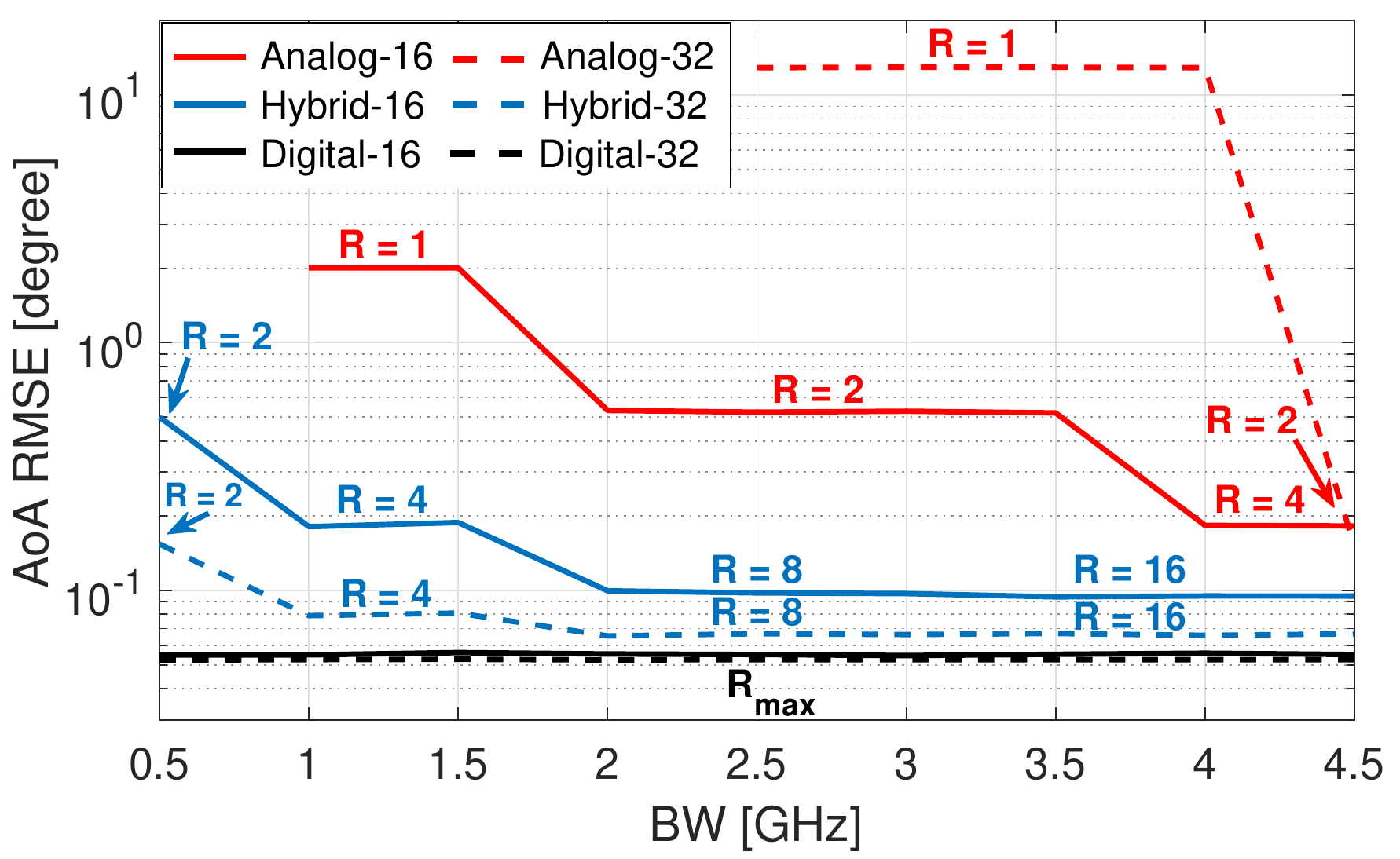}
    \end{center}
    \vspace{-4mm}
    \caption{Beam training performance comparison of the three considered architectures and the interplay of $R$, $N_{\R}$, and $\BW$.}
    \vspace{-4mm}
    \label{fig:interplay}
\end{figure}
We study the beam training performance of different architectures in terms of \ac{AoA} estimation accuracy, assuming that $R$ is constrained to be maximal power of 2. We consider a system with carrier frequency $\fc=$ \SI{60}{\giga\hertz}, bandwidth values in the range $0.5~\text{GHz}\leq \BW \leq 4.5~\text{GHz}$, and $M_{\tot}=4096$ subcarriers for any bandwidth. The transmitter array size is $N_{\T}=128$, while the receive array size can take values $N_{\R} = \{16, 32\}$. There are $N_{\text{r}}=4$ antennas in each sub-array in hybrid \ac{TTD} architecture, regardless of the total number of antennas. The number of probed directions in beam training is assumed to be $D=2N_{\R}$ and the dictionary size is $Q=1024$. The channel consists of $L=3$ clusters, where one is $10$~dB stronger than the other two. Fading is simulated by 20 rays within each cluster with up to \SI{10}{\nano\second} spread. There is no intra-cluster angular spread. Pre-beamforming \ac{SNR} is defined as $\SNR \triangleq \sum_{l=1}^{L}  \sigma_l^2/\sigmaN^2$, and it is assume to be $\SNR=$ \SI{-20}{\decibel}.

In Fig.~\ref{fig:interplay}, we present the results for the beam training performance and the interplay of the considered parameters. In both cases $N_{\R}=16$ and $N_{\R}=32$, the analog \ac{TTD} array architecture has the highest \ac{RMSE} of \ac{AoA} estimation due to low achievable diversity order $R$. As discussed earlier, analog arrays have large delay range requirements, and thus better estimation accuracy (equivalently, higher $R$) requires larger $\BW$. Similarly, increasing the array size $N_{\R}$ can have a positive effect on the performance. However, if $\BW$ is not large enough and there is no diversity $(R=1)$, larger arrays do not improve the estimation accuracy in frequency-selective channels. The analog arrays do not have the results for the values of $\BW$ for which the proposed single-shot beam training cannot be realized $(R<1)$. In hybrid \ac{TTD} arrays, higher diversity orders can be utilized since $N_{\text{r}} < N_{\R}$, which leads to better estimation accuracy compared to analog arrays. Increase in the number of antenna elements does not change achievable $R$ in hybrid arrays since we assume that $N_{\text{r}}=4$ remains constant. It does, however, improve the estimation accuracy of hybrid arrays, which approaches the sub-degree performance of fully digital arrays. Since $R$ can be maximized through \ac{DSP} in digital arrays, their performance is independent of $\BW$. The floor of the \ac{AoA} \ac{RMSE} is determined by the dictionary size $Q=1024$. Based on described results in Fig.~\ref{fig:interplay}, one can predict the diversity order $R$ and beam training performance for any considered array architecture, given the system parameters $\BW$, $N_{\R}$, and $\Tcmax$.

\begin{figure}
    \begin{center}
        \includegraphics[width=0.48\textwidth]{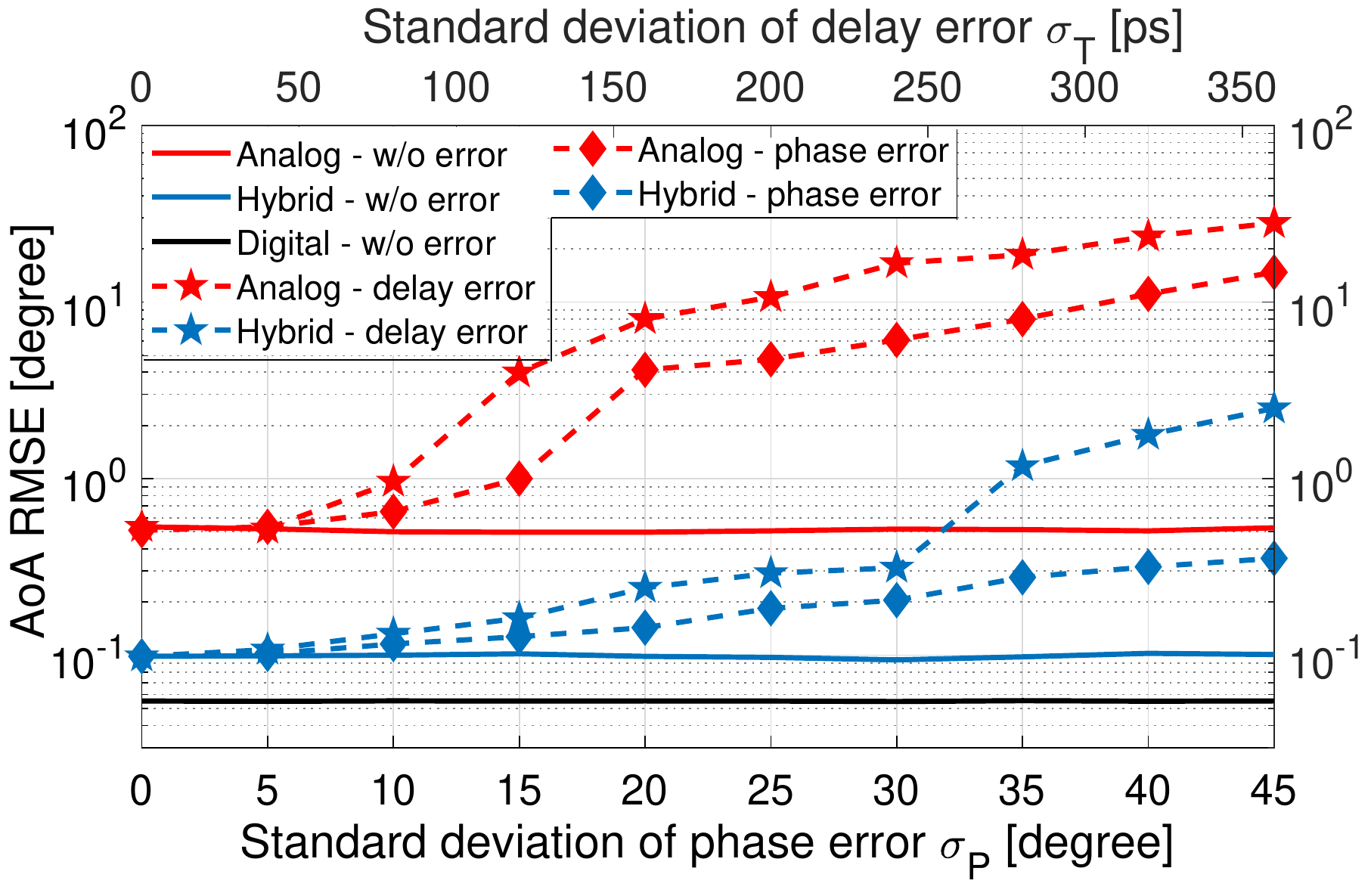}
    \end{center}
    \vspace{-4mm}
    \caption{Beam training performance comparison of the three considered architectures under the distorted delay taps $\tilde{\tau}_n\sim\mathcal{N}\left(\tau_n,\sigma_{\text{T}}^2\right),~\forall n$, and phase taps $\tilde{\phi}_n\sim\mathcal{N}\left(\phi_n,\sigma_{\text{P}}^2\right),~\forall n$. The curves with the delay error (dashed with stars) and phase error (dashed with diamonds) are associated with the upper and lower x-axis, respectively.}
    \vspace{-4mm}
    \label{fig:delay_phase_error}
\end{figure}

\subsection{Impact of \ac{TTD} Hardware Impairments on Beam Training}

Next, we study the impact of practical \ac{TTD} hardware impairments and \ac{ADC} quantization errors on beam training in all considered architectures. Here we keep \ac{AoA} \ac{RMSE} as the performance metric and use the same system parameters as in the previous subsection. We consider a specific case with $N_{\R} = 16$ and $\BW = 2$ GHz.

In Fig.~\ref{fig:delay_phase_error}, we study the beam training performance under the phase and delay errors. Unlike analog and hybrid \ac{TTD} arrays, fully digital array is not sensitive to these hardware impairments and we include its performance with the maximum $R$ as the benchmark. With the considered system parameters, analog \ac{TTD} array has the diversity order $R=2$, which limits its angle estimation accuracy and robustness to hardware errors. We can see that the beam training algorithm can tolerate phase errors with the standard deviation of up to $\sigma_{\text{P}}=$ \SI{15}{\degree} and delay errors with the standard deviation of up to $\sigma_{\text{T}}=$ \SI{75}{\pico\second}. Hybrid \ac{TTD} array achieves a lower estimation accuracy and greater robustness to delay and phase errors than analog \ac{TTD} array since it leverages the diversity order $R=8$ in beam training. It can tolerate large phase errors and delay errors with the standard deviation larger than $\sigma_{\text{T}}=$ \SI{200}{\pico\second}. It is worth noting that the delay errors in hybrid arrays are independent of the reduced delay taps in the corresponding \ac{TTD} elements.

\begin{figure}[t]
    \begin{center}
        \includegraphics[width=0.48\textwidth]{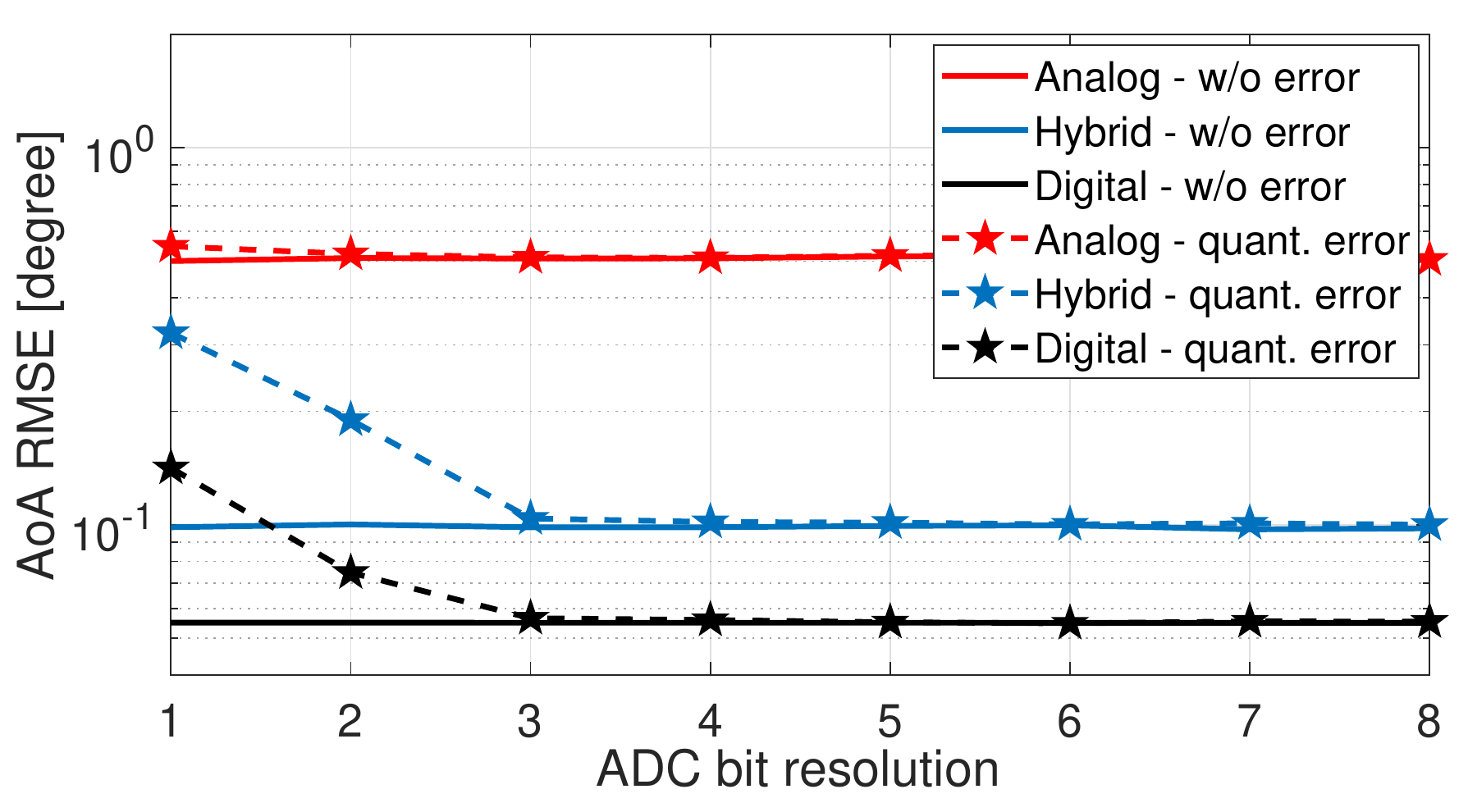}
    \end{center}
    \vspace{-4mm}
    \caption{Beam training performance comparison of the three considered architectures under different \ac{ADC} resolutions.}
    \vspace{-4mm}
    \label{fig:quantization_error}
\end{figure}
In Fig.~\ref{fig:quantization_error}, we present how finite \ac{ADC} resolution affects the beam training performance with different array architectures. For fair comparison, we assume that the \ac{AGC} outputs a unit-variance signal in all architectures. We can observe that the \ac{AoA} estimation accuracy of the analog \ac{TTD} array with a single \ac{RF}-chain is marginally affected by low \ac{ADC} resolution. On the other hand, low resolution \ac{ADC}s have a noticeable impact on beam training with the hybrid \ac{TTD} and fully digital arrays, as combined quantization errors from different \ac{RF}-chains deteriorate the estimation accuracy. We note, however, that the deteriorated accuracy is still within the sub-degree range and lower than that of the analog array. Our results indicate that practical \ac{mmW} and sub-THz transceivers may require \ac{ADC}s with only a few bits of resolution for effective beam training. For example, with only 3-bit resolution, the performance loss in negligible in any array. Low-resolution \ac{ADC}s have a positive impact on the overall power efficiency of the considered \ac{TTD} architectures, as discussed in the next section.

%
%

\section{Power Analysis of TTD Architectures} 
\label{sec:poweranal}

This section presents power analysis of the analog and hybrid TTD arrays comparing it with a digital array for the proposed mmW beam training algorithm in Section~\ref{sec:algorithm}. We will estimate the power consumption of the baseband components in the signal chain in Fig.~\ref{fig:analog}, Fig.~\ref{fig:hybrid}, and Fig.~\ref{fig:digital} for the analog, hybrid, and digital arrays assuming the mmW front-end consumes the same power in all the three array architectures. The only exception to this assumption in the front-ends of the three array architectures is the phase-shifter. In the analog/hybrid TTD array , the phase shifter precedes the downconverting mixer whereas for the digital array it can be implemented after the ADC. For the sake of comparison in this work, we consider the digital phase-shifter power equivalent to that of the RF/LO phase-shifter which will be assumed to have a complete passive implementation \cite{elkholy2018}.  The estimation methodology for the remaining components of the hybrid and the digital arrays follows that of the analog TTD array as described in the next subsections. For each component, we also have provided an example based on Table~\ref{tab:ttdspec}.


\subsection{Power Consumption of Analog/Hybrid \ac{TTD} Array}
Referring Fig.~\ref{fig:analog} (Fig.~\ref{fig:hybrid}), this subsection will estimate the power consumption of the ADC, AGC, SCA, and the time-interleaving blocks that differentiates the power consumption in the analog (hybrid) arrays.  

\subsubsection{Analog-to-Digital Converter (ADC)} We estimate the \ac{ADC} power consumption using \ac{FoM} derived from recent works \cite{oh2019,zhu2018,Chan2017} on low-resolution high-speed flash \ac{ADC}s  (different ADC configuration can be selected when considering efficiency). Using the \ac{FoM} of the state-of-the-art flash \ac{ADC}s from Table~\ref{tab:FoMADC}, we take the average \ac{FoM} of $96.1$fJ/c-s for our estimation. For a 3-bit ENOB, $\fs$=4GHz and a \ac{FoM} of $96.1$fJ/c-s, the estimated power is thus $3.07$mW. 

In addition to the ADC power consumption, we also estimate the deserializer power that is needed to interface the high-speed ADCs with the backend DSP. Though insignificant for analog and hybrid arrays, it will be an important contributor for digital arrays. We consider here the \ac{DSP} operating at 1GHz and estimate the deserializer power consumption. From \cite{jung2013}, excluding the power of clock generator, the scaled deserializer power for one unit ($\PDESo$) is found to be $0.512$~mW $(=3.2\times4/25)$ which yields 1.5mW and 6mW of power consumption in analog and hybrid array respectively. 

\begin{table}[t]
\vspace{0mm}
\centering
\caption{\small State-of-the-art low-resolution GHz ADCs.} 
\label{tab:FoMADC}
\vspace{-2mm}
{   
    \footnotesize { 
    \begin{tabular}{|c||c|c|c|} 
    \hline
        Parameters & \cite{oh2019} & \cite{zhu2018} & \cite{Chan2017}  \\
        \hline
        \hline
        Sampling Rate (f\textsubscript{s})(MHz)  &  2500 & 2000 & 5 \\
        \hline
        ENOB (bit)   &  6 & 7.93 & 4.06  \\
        \hline 
        Power ($\mu$W)  &  7500 &  21000 & 78000 \\
        \hline
        FoM (fJ/c-s)  &  74.7 & 119 & 94.6 \\ 
        \hline 
        Technology (nm)  &  65 & 65 & 65 \\ 
        \hline 
    \end{tabular} \\
}
}
\vspace{1mm}
\end{table}




\subsubsection{Switched-Capacitor Array (SCA)} The SCA power consumption is dominated mostly by the feedback \ac{OTA}. We estimate the OTA power consumption for an analog array similar to the method in \cite{ghaderi2019}. The DC gain (A0) and the unity-gain bandwidth ($\omega_u$) requirements of the OTA used in the \ac{SCA} are found to be: 
\begin{align*}
    \omega_{\text{u}} = 2 \ln(2) N_{\R} (x+1) \fs 
\end{align*}
where $x$ is the \ac{ADC} resolution and $\fs$ is the \ac{ADC} sampling frequency.

The normalized unity-gain bandwidth ($\omega_{\text{u0}}$) per unit sampling frequency can be written as: $\omega_{\text{u0}}= 2 \ln(2) (x+1)$. For a 3-bit \ac{ADC} (referring Fig.~\ref{fig:quantization_error}), the normalized unity-gain frequency $\omega_{\text{u0}}= 2 \ln(2) (3+1)=$\SI{5.54}{\hertz}. Neglecting parasitics, second order effects, and considering a two-stage internally compensated OTA, the transconductance of this OTA can be designed to be linearly dependent to the DC current. As a result, the DC gain of the OTA is independent of its DC current and power consumption $P_{\text{OTA}}$. At the same time, $\omega_{\text{u}}$ is a linear function of the OTA transconductance, and thus varies proportionally to $P_{\text{OTA}}$. Given these assumptions, the minimum requirement on the OTA $\omega_{\text{u}}$ results in linear dependency of $P_{\text{OTA}}$ to the product of the number of antennas and sampling frequency, as shown below \cite{ghaderi2019}:
\begin{align*}
    \label{eq:otapower}
    P_{\text{OTA}} \approx P_{\text{OTAo}} N_{\R} \fs 
\end{align*}
where $P_{\text{OTAo}}$ is the power consumption of an OTA designed for a single-element array with unit sampling frequency (1Hz). Solving for a $60^{\circ}$ \ac{PM} requirement puts $C_{\text{c}}$ close to $0.22$~pF yielding $g_{\text{mn}}=0.22~\text{pF} \times 5.54=1.2188~\text{ps}$. Assuming $g_\text{m}/I_\text{D} =15$, the unit current can be obtained as $I_{\text{Dn}}=8.1253 \times 10^{-14}$~A. For a $60^{\circ}$ \ac{PM}, the $g_\text{mn}$ for the second stage is around 10 times of the first stage and we further assume the same $g_\text{m}/I_\text{D}$ ratio. The total current is thus $(2+10) I_{\text{Dn}}= 9.7504\times10^{-13}$A. Assume a 1V supply, the $P_{\text{OTAo}}$ can be estimated as $9.7504\times 10^{-13}$W. For the 16-antenna array and $\mathrm{f_s=4GHz}$ (Table~\ref{tab:ttdspec}), the estimated power consumption is thus $62.403$ mW. Note that the hybrid TTD array relaxes the OTA power consumption per sub-array where $P_{\text{OTA}}$ is scaled by $N_\text{R}/N_\text{H}$.  The same power consumption estimation however applies to a digital array without any relaxation. 


The power consumption of the \ac{AGC} can also be estimated using, $\PAGC$. Assuming $\PAGC$ consumes the same power as the OTA, the estimated total $\PAGC$ is also equal to $3.9$mW for analog arrays and $15.6$mW for the hybrid array following the design specifications in Table~\ref{tab:ttdspec}.

\subsubsection{Time-Interleaver} 
The power consumption for the time interleaver can be estimated as \cite{razavi2013}: 
\begin{align*} 
    \PTINW=\fs/N_{\I} \times N_{\I} \times (\Csw/N_{\I} + \Cint) \times \VDD^2
\end{align*}
where $\Csw$ is the switch capacitance and $\Cint$ is the interconnect parasitic capacitance. For a sampling frequency of 4GHz and 1V supply, $\Csw=2.5$pF, $\Cint=0.6$pF \cite{razavi2013}, 31 levels of time interleaving in analog array, and \textcolor{black}{7 levels of interleaving in a hybrid array}, the estimated power consumption of the time interleaver is $2.7$mW and $3.8$mW for the analog and digital hybrid arrays respectively.

\subsection{Power Consumption of Digital Array} 
The estimated power consumption for digital TTD array can be derived following a similar approach to the analog arrays with the important consideration that the proposed beam training algorithm will require only integer delays at the ADC sampling frequency. For operation in communication mode, fractional-rate samplers will be needed as detailed in \cite{jang2019a}. In addition to the same number of \ac{ADC}s, \ac{AGC}s and filters as in an analog TTD array, the digital array consumes higher power at the ADC-DSP interface primarily due to the need for de-serializing the high-speed ADC output. For example, with 16-elements and 3-bit per \ac{ADC}, the estimated power consumption of the deserializer will be $24.6$mW.
\begin{table}[t]
\vspace{0mm}
\centering
\caption{\small Power estimation methodology for TTD arrays.} 
\label{tab:PrwDsgMet}
\vspace{-2mm}
{   
    \footnotesize { 
    \begin{tabular}{|c||c|c|c|} 
    \hline
        $\#$ Components & Analog & Hybrid & Digital \\
        \hline
        \hline
        ADC  &  1 & $N_{\text{H}}$ & $N_{\R}$ \\
        \hline 
        SCA/AGC  &  1 &  $N_{\text{H}}$ & $N_{\R}$ \\
        \hline 
        \hline
        $\PSCA$ & $\PSCAo N_{\R} \fs$ & $\PSCAo {N_{\R}}/{N_{\text{H}}} \fs$  & $\PSCAo N_{\R} \fs$  \\ 
        \hline
        $P_{\text{ADC}_{x-\text{bit}}}$ & \multicolumn{3}{c|}{FoM based estimation}  \\ 
        \hline
        $\PAGC$ &  $\PSCAo \fs$ & $\PSCAo N_{\text{H}} \fs$  & $\PSCAo N_{\R}  \fs$ \\
        \hline
        $\PDeSer$ &  $\PDESo  \fs  x $ & $\PDESo N_{\text{H}} x $  & $\PDESo N_{\R}  x$ \\
        \hline
    \end{tabular} \\
}
}
\vspace{1mm}
\end{table}
\vspace{0mm}


\subsection{Comparison of Estimated Array Power Consumption} 
\begin{figure}
    \begin{center}
        \includegraphics[width=0.4\textwidth]{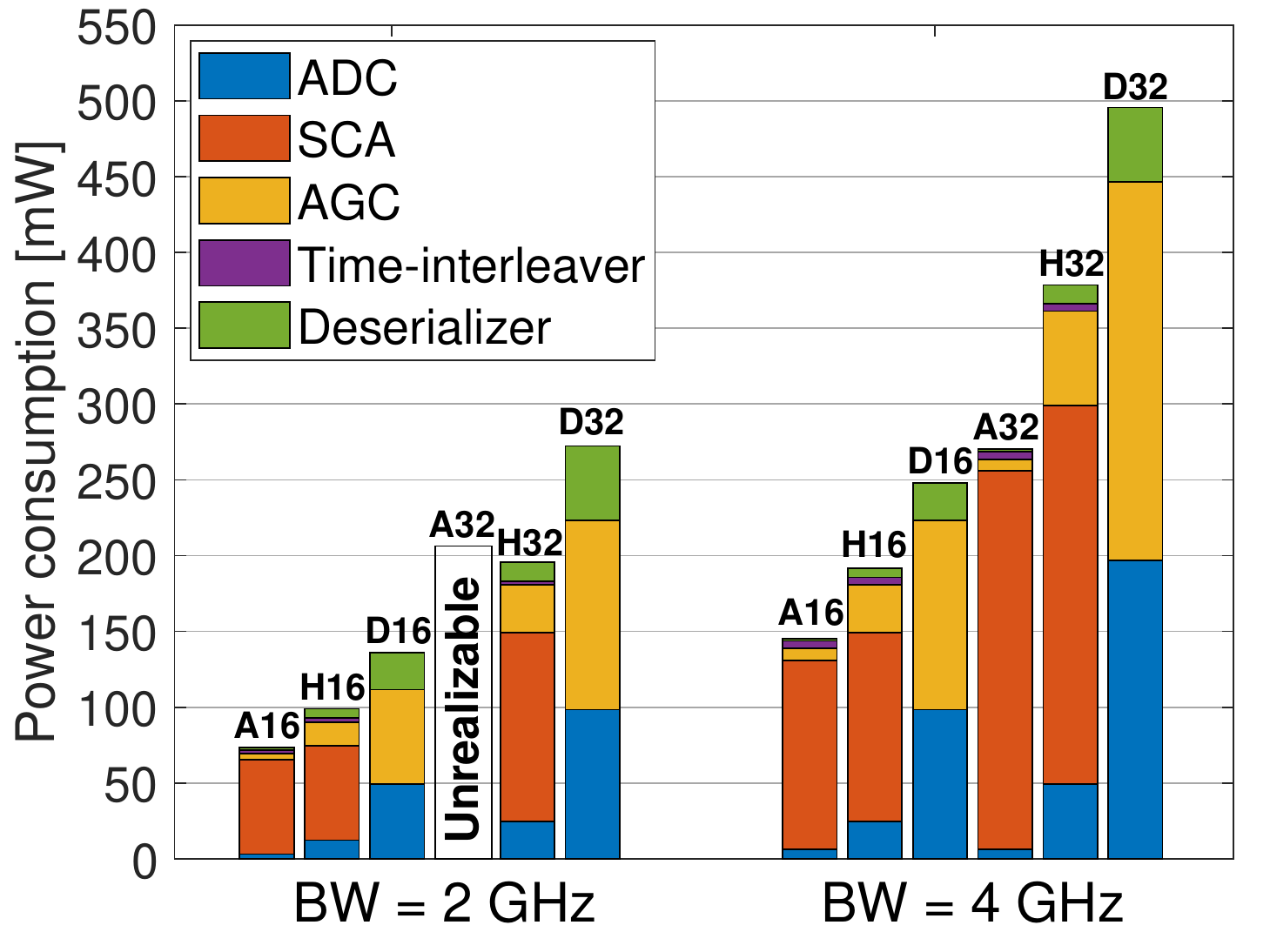}
    \end{center}
    \vspace{-4mm}
    \caption{Comparison of analog (A), hybrid (H), and digital (D) architectures in terms of power consumption for $N_{\R}=\{16, 32\}$ and $\BW = \{2, 4 \}$GHz.}
    \vspace{-4mm}
    \label{fig:spef_power}
\end{figure}

Table IV summarizes the required number of components and power consumption in the analog, hybrid, and digital arrays based on the architectures in Fig.~\ref{fig:analog}, Fig.~\ref{fig:hybrid}, and Fig.~\ref{fig:digital}, respectively. Fig.~\ref{fig:spef_power} illustrates the introduced power estimation methodology with a breakdown of individual components for the analog and hybrid TTD arrays and also the benchmark digital array. The estimated power consumption for each component block is described in the previous subsections for each array architecture. The analog array provides high energy efficiency as compared to the hybrid TTD array and digital arrays. However, the increasing bandwidth as well as the number of elements requires larger unity-gain bandwidths OTAs which increases design complexity for higher diversity orders. The need for higher unity-gain bandwidths is further constrained with increasing number of feedback to the virtual ground of the OTA. Hybrid arrays are thus the most optimal choice to meet feasible delay-bandwidth products. Future work will investigate design of analog (hybrid) arrays with higher number of antenna elements per sub-array  using passive SCA that leverages the reasonably lower resolutions ($\approx3$-bit) required by the proposed beam training algorithm. 

%
%

\section{Conclusions and Future Work} 
\label{sec:concl}

This work introduced and analyzed two \ac{TTD} architectures with large delay-bandwidth product baseband delay elements as potential candidates for \ac{mmW} beam training. We demonstrated that a high \ac{AoA} estimation accuracy can be achieved with both proposed \ac{TTD} architectures using a power measurement based beam training scheme, which requires only one wideband training pilot. The dependency of the codebook design and beam training performance on system parameters, including the bandwidth, number of antenna elements, and maximum \ac{TTD} delay compensation, was analyzed and numerically evaluated in a practical multipath fading channel. Detailed analysis of the angle estimation accuracy, robustness to hardware impairments, and power consumption, revealed the trade-offs between the proposed \ac{TTD} architectures when benchmarked against the digital array. The analog \ac{TTD} array consumes 66\% less power than the digital array, but it achieves a higher angle estimation error. The hybrid \ac{TTD} array has a comparable beam training performance and 25\% lower power consumption than the digital array. The results on how power consumption scales with the key system parameters, including the bandwidth and array size, provided an insight into the beam training design for future \ac{mmW} and sub-THz systems. Future work will include array implementations supporting larger delay-bandwidth products for arrays with higher number of antenna elements, as well as channel estimation and identification of multiple \ac{AoA}s in interference-limited networks.


%

\appendices

\section{Derivation of Expected Powers in $D$ Directions}
\label{sec:alg_appendix}

We consider a frequency-selective multipath channel with $L$ clusters and corresponding gains modeled as $G_l\sim \mathcal{CN}\left(0, \sigma_l^2 \right)$. The channel gains are assumed to be independent over different clusters and frequency sub-bands. The frequency domain channel model can be approximated as
\begin{equation}
    \label{eq:ch_approx}
    \mathbf{H} = \mathbf{A}_{\R} \mathbf{\Lambda} \mathbf{A}_{\T}^{\hermitian},
\end{equation}
where $\mathbf{A}_{\R}\in\mathbb{C}^{N_{\R}\times Q}$ and $\mathbf{A}_{\T}\in\mathbb{C}^{N_{\T}\times Q}$ contain $Q$ array responses $\mathbf{a}_{\R}(\xi_q)$ and $\mathbf{a}_{\T}(\xi_q)$ that correspond to $Q$ uniformly spaced angles $\xi_q,~q=1,...,Q$, in the range $\left(-\pi/2,\pi/2\right)$. The elements of array responses are defined as $[\mathbf{a}_{\R}(\theta)]_n = N^{-1/2}_{\R} \mathrm{exp}({-j(n-1)\pi\sin({\theta)}}),~n=1,...,N_{\R}$, and $[\mathbf{a}_{\T}(\theta)]_n = N^{-1/2}_{\T}\mathrm{exp}({-j(n-1)\pi\sin{(\theta)}}),~n=1,...,N_{\T}$. The square matrix $\mathbf{\Lambda}\in\mathbb{C}^{Q\times Q}$ has only $L$ non-zero elements that correspond to the gains $G_l,~\forall l$. Commonly, $Q\gg L$ and the approximation error in (\ref{eq:ch_approx}) can be neglected. 

With the codebook design described in \Cref{sec:algorithm}, the received signal in any sounded direction $d$ can be considered as a zero-mean complex Gaussian random variable and expressed as
\begin{equation}
\label{eq:received_signal_direction}
    Y_d = \mathbf{f}_d^{\hermitian} \mathbf{H} \mathbf{v} + \mathbf{f}_d^{\hermitian}\mathbf{n},
\end{equation}
where $\mathbf{n} \sim  \mathcal{CN}\left(0, \sigmaN^2 \mathbf{I}_{\R} \right)$ is white Gaussian noise. The realizations of (\ref{eq:received_signal_direction}) are received symbols $Y[m],~m\in\mathcal{M}_d$. The expected received signal power in direction $d$ is $p_d = \mathbb{E}\left[ \abs{Y_d}^2 \right] = \mathbb{E}[ (\mathbf{f}_d^{\hermitian} \mathbf{H} \mathbf{v}M^{-1/2} + \mathbf{f}_d^{\hermitian} \mathbf{n} )^{\hermitian} (\mathbf{f}_d^{\hermitian} \mathbf{H} \mathbf{v}M^{-1/2} + \mathbf{f}_d^{\hermitian} \mathbf{n} ) ]$. Based on the channel model in (\ref{eq:ch_approx}), it can be shown that
\begin{align}
    \label{eq:expected_power_initial}
    p_d &= M^{-1} \mathbb{E} \left[ \mathbf{v}^{\hermitian}\mathbf{A}_{\T} \mathbf{\Lambda}^* \mathbf{A}_{\R}^{\hermitian}\mathbf{f}_d\mathbf{f}_d^{\hermitian} \mathbf{A}_{\R} \mathbf{\Lambda} \mathbf{A}_{\T}^{\hermitian} \mathbf{v} \right] + \mathbb{E} \left[ \mathbf{n}^{\hermitian}\mathbf{f}_d\mathbf{f}_d^{\hermitian}\mathbf{n} \right].
\end{align}

We apply the trace operator $\text{Tr}()$ to (\ref{eq:expected_power_initial}) and exploit its linearity and cyclic property to obtain
\begin{align*}
    p_d &= M^{-1} \mathbb{E} \left[ \text{Tr}\left( \mathbf{\Lambda} \mathbf{A}_{\T}^{\hermitian} \mathbf{v} \mathbf{v}^{\hermitian}\mathbf{A}_{\T} \mathbf{\Lambda}^* \mathbf{A}_{\R}^{\hermitian}\mathbf{f}_d\mathbf{f}_d^{\hermitian} \mathbf{A}_{\R} \right) \right] + N_{\R}\sigmaN^2\\
    &= \text{Tr}\left( \mathbf{G} \mathbf{A}_{\R}^{\hermitian} \mathbf{f}_d \mathbf{f}_d^{\hermitian} \mathbf{A}_{\R}\right) + N_{\R}\sigmaN^2. \numberthis \label{eq:expected_power_trace}
\end{align*}
where $\mathbf{G} =  \mathbb{E} \left[ \mathbf{\Lambda} \mathbf{A}_{\T}^{\hermitian} \mathbf{v} \mathbf{v}^{\hermitian}\mathbf{A}_{\T} \mathbf{\Lambda}^* \right]$. Since $\mathbf{\Lambda}$ and $\mathbf{\Lambda}^*$ are sparse matrices, $\left[ \mathbf{\Lambda} \mathbf{A}_{\T}^{\hermitian} \mathbf{v} \mathbf{v}^{\hermitian}\mathbf{A}_{\T} \mathbf{\Lambda}^* \right]$  yields another sparse $Q\times Q$ matrix with $L^2$ non-zero elements. There are $L$ non-zero elements of the form $\abs{G_l}^2\abs{\mathbf{a}_{\T}^{\hermitian}(\theta^{(T)}_l) \mathbf{v}}^2,~\forall l$, on the main diagonal. The $L(L-1)$ off-diagonal elements are cross terms $G_{l_1}G_{l_2}^*\mathbf{a}_{\T}^{\hermitian}(\theta^{(T)}_{l_1}) \mathbf{v} \mathbf{v}^{\hermitian}\mathbf{a}_{\T}(\theta^{(T)}_{l_2}),~\forall l_1,l_2$. Thus, $\mathbf{G}$ is a diagonal matrix with $L$ non-zero elements $ \sigma_l^2\abs{\mathbf{a}_{\T}^{\hermitian}(\theta^{(T)}_l) \mathbf{v}}^2,~\forall l$, since $\mathbb{E}\left[G_{l_1}G^*_{l_2} \right]=0,~\forall l_1 \neq l_2$, and  $\mathbb{E}\left[\abs{G_l}^2 \right]=\sigma_l^2,~\forall l$. The product of $\mathbf{G}$ and the matrix of the \ac{UE} \ac{BF} gains $\mathbf{A}_{\R}^{\hermitian} \mathbf{f}_d \mathbf{f}_d^{\hermitian} \mathbf{A}_{\R}$ is a $Q\times Q$ matrix whose diagonal elements are equal to $\abs{\mathbf{f}_d^{\hermitian}\mathbf{a}_{\R}(\xi_q)}^2 [\mathbf{G}]_{q,q}$, so (\ref{eq:expected_power_trace}) becomes
\begin{align}
    \label{eq:expected_power_one_dir}
    p_d = \mathbf{b}_d^T \mathbf{g} + N_{\R}\sigmaN^2
\end{align}
where $\mathbf{b}_d = \left[ \abs{\mathbf{f}_d^{\hermitian}\mathbf{a}_{\R}(\xi_1)}^2, \abs{\mathbf{f}_d^{\hermitian}\mathbf{a}_{\R}(\xi_2)}^2, ..., \abs{\mathbf{f}_d^{\hermitian}\mathbf{a}_{\R}(\xi_Q)}^2 \right]^T$ and $\mathbf{g}=\text{diag}(\mathbf{G})$. By vectorizing the result in $(\ref{eq:expected_power_one_dir})$, we obtain
\begin{align}
    \label{eq:expected_power_matrix_appendix}
    \mathbf{p} = \mathbf{B}\mathbf{g} + N_{\R}\sigmaN^2\mathbf{1},
\end{align}
where $\mathbf{p}=\left[p_1,p_2,...,p_D\right]^T$ and $\mathbf{B}=\left[\mathbf{b}_1,\mathbf{b}_2,...,\mathbf{b}_D \right]^T$. Since the \ac{BS} provides a large \ac{BF} gain with the fixed precoder $\mathbf{v}$, we can assume that receiver array sees only one spatially filtered dominant cluster, e.g., the first one. Consequently, there is only one non-zero element in $\mathbf{g}$ equal to $\abs{ \mathbf{a}_{\T}^{\hermitian}(\theta^{(\T)}_{1}) \mathbf{v} }^2 \sigma_1^2$.

\ifCLASSOPTIONcaptionsoff
  \newpage
\fi



%
%
%

\bibliographystyle{IEEEtran}
\bibliography{IEEEabrv,references,ref_SG}

\end{document}